# Althea: The Fact-Checking–Metalearning Tradeoff in AI-Assisted Verification

SVETLANA CHURINA[2], KOKIL JAIDKA[4], ANAB MAULANA BARIK[3], HARSHIT ANEJA, CAI YANG[4], INSYIRAH BINTE IMAM MUJTAHID, WYNNE HSU[2,3], and MONG LI LEE[2], Centre for Trusted Internet & Community (CTIC), [2] Institute of Data Science, [3] School of Computing, [4]Department of Communications and New Media, National University of Singapore (NUS), Singapore

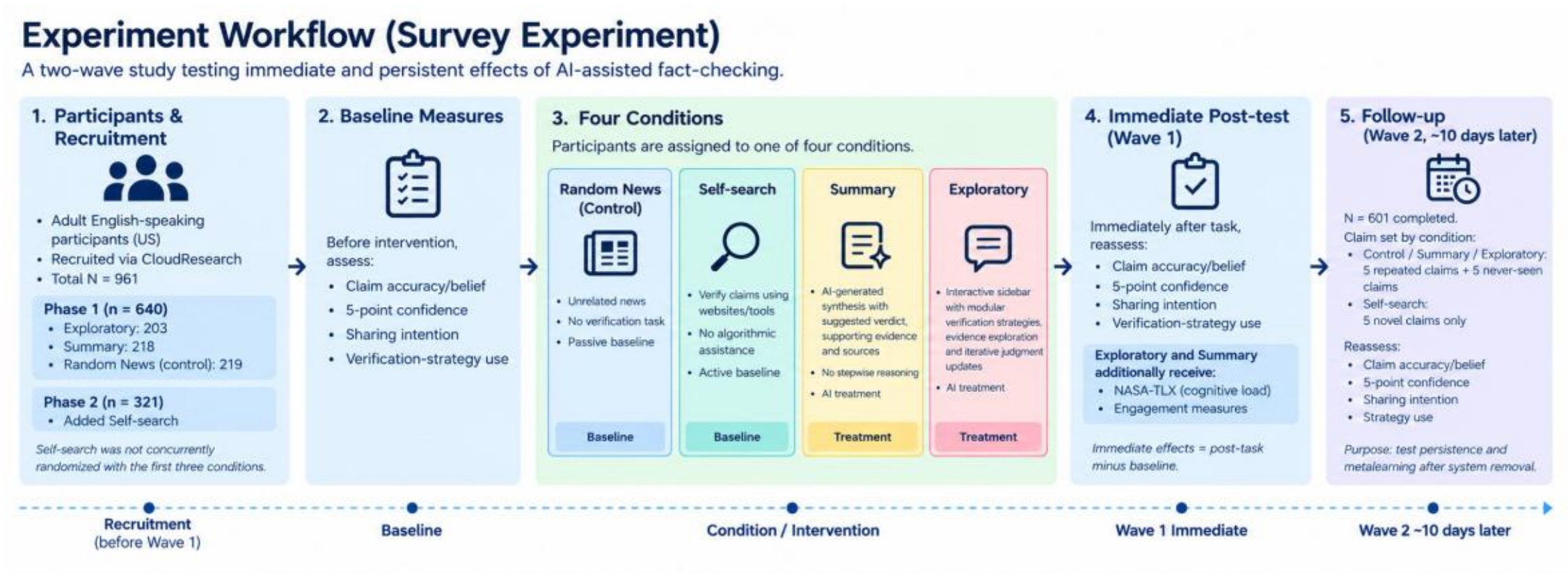

Fig. 1. Workflow of the survey experiment (Study 2). Participants ($N$ = 961; Exploratory, Summary, and Random News randomized in phase one, Self-search added in phase two) complete baseline measures, are assigned to one of four conditions that vary how much of the verification procedure the system performs for them, and are reassessed immediately (Wave 1) and ten days later with the system removed (Wave 2), on claims repeated from Wave 1 and on claims no condition had seen. Both AI-assisted treatments produced the largest immediate gains in accuracy and confidence, but on the never-seen claims at Wave 2 they were indistinguishable from Random News, while Self-search, which received no system-authored verdict, retained a large advantage: the conditions that most improve immediate accuracy are the least likely to produce metalearning.

Fact-checking systems must be scalable and epistemically trustworthy. We introduce *Althea*, a retrieval-augmented system for user-driven claim evaluation that matches standard pipelines on the AVeriTeC dataset and shows a particular improvement on challenging cases with mixed evidence. A longitudinal survey experiment ($N$ = 961) treats a ten-day follow-up as a *fading test*: after modeling a verification procedure, we remove the system and ask whether users reproduce it unaided, testing *metalearning* rather than one-time accuracy. We compare two AI-assisted treatments, *Exploratory* (guided reasoning) and *Summary* (synthesized verdicts), against two baselines, unrelated news and *Self-search*. The treatments yield the strongest immediate accuracy and confidence gains but do not survive the fading test: on unseen claims they perform no better than news, while Self-search, with no procedure to fade, retains a

Authors' Contact Information: Svetlana Churina[2] , churinas@nus.edu.sg; Kokil Jaidka[4] , jaidka@nus.edu.sg; Anab Maulana Barik[3] , anabmaulana@u.nus.edu; Harshit Aneja, contactaneja@gmail.com; Cai Yang[4] , cai.yang@u.nus.edu; Insyirah Binte Imam Mujtahid, insyirah@nus.edu.sg; Wynne Hsu[2,3] , whsu@comp.nus.edu.sg; Mong Li Lee[2] , leeml@comp.nus.edu.sg, Centre for Trusted Internet & Community (CTIC), and [2] Institute of Data Science, and [3] School of Computing, and [4]Department of Communications and New Media, National University of Singapore (NUS), Singapore, Singapore.

large advantage. This reveals a fact-checking–metalearning tradeoff: conditions that most improve immediate accuracy are least likely to produce metalearning, cautioning against treating AI-delivered verdicts as a source of durable literacy gains.

## 1 Introduction

Chatbots are increasingly used to perform tasks once handled by search engines and static fact-check pages: answering questions, summarizing what they retrieve, and resolving disputed claims in real time. Half of U.S. adults now use an AI chatbot and a quarter use one daily; searching for information is the most common use, reported by 42% of adults, up from a third two years earlier [22], and checking whether something one has read is true is among the uses people report [13]. As large language models become a default interface for information-seeking, what a chatbot-mediated correction actually leaves behind, once the chatbot is no longer there, is an urgent design and policy question, and one that the accuracy of the correction itself does not answer.

Fact-checking requires domain knowledge, interpretive judgment, and an ability to evaluate conflicting evidence under uncertainty. It is cognitively demanding [25], and in online environments, it must often be done rapidly in the face of high-volume and emotionally charged content [54]. While automated systems have made progress in scaling verification [1, 2, 18, 38, 49], and others have improved transparency through explanation and attribution [14, 16, 41], these approaches often treat human users as passive recipients of system outputs, as opposed to active reasoners engaged in evaluation.

Three observations from prior work motivate our design. First, verification is a reasoning problem and not merely an information-retrieval problem: even when relevant information is accessible, users struggle to identify which evidence is diagnostic or to reconcile conflicting sources, to the point that asking people to search online to evaluate a claim can *increase* their belief in false ones [4]. Effective verification depends on how information is structured and how users are guided through reasoning [19, 29, 59], and performance remains modest without such support [5, 43, 51]. Whether users can integrate what they find into their own judgments, and remain the source of those judgments rather than recipients of system-generated verdicts, is what we call *epistemic agency* [20]. Second, scaffolding improves verification, but its effects on evaluation accuracy are not uniform: explicitly teaching lateral-reading strategies improves both strategy use and accuracy [10, 11], yet even after such instruction, learners continue to rely on weak credibility cues and need further support in reasoning about source expertise and trustworthiness [37].

Third, improved performance with support does not necessarily translate into independent verification ability: Rani et al. [42] found that persuasive AI news-verification dialogue produced immediate accuracy gains but a significant decline in unassisted detection four weeks later. Whether withholding the verdict altogether, and not merely changing how it is delivered, would avoid that decay is the question we address directly.

Measuring persistence by re-testing the same claims ten days later carries a risk of its own: repeated exposure to a claim, independent of any reasoning about it, is known to shift how it is judged. The illusory-truth-effect literature shows that mere repetition increases perceived accuracy regardless of actual veracity [24], and this operates on misinformation specifically, prior exposure to a false claim increases its later perceived accuracy [39]. A persistence measure built only from repeated items cannot separate durable reasoning from this familiarity effect, and for real misinformation the same mechanism could inflate confidence in a false claim as readily as a true one.

To this end, we introduce Althea, a retrieval-augmented fact-checking system, and test it as two AI-assisted treatments against two baselines. The **Exploratory** treatment scaffolds reasoning through modular evidence exploration without providing a verdict (high in encouraging learning, moderate in authorship). The **Summary** treatment provides a synthesized verdict with supporting evidence (high correct-information delivery, low authorship). A walkthrough of the Althea interface, and both treatment conditions, is provided as a video included in the supplementary materials of this submission.

We evaluate the efficacy of these treatments against two baselines. **Self-search** is an active baseline: participants receive only procedural guidance and perform all evidence retrieval and synthesis themselves, with no algorithmic assistance, so that whatever the treatments deliver can be measured against what unaided effort on the identical task already achieves. **Random News** is a passive baseline with no verification task at all. Together they let us ask one question, *does AI assistance in fact-checking teach people to fact-check, or only to accept its verdicts while it is present?*, in two parts:

- **RQ1:** Do the AI-assisted treatments produce larger immediate gains in fact-checking accuracy and confidence than the two baselines?
- **RQ2:** Do the treatments retain this advantage over both baselines at 10-day follow-up, or does their benefit fail to survive removal of the system?

Comparing these conditions across immediate and delayed measurements separates two things a single AI-assisted correction can leave behind. It can produce *object-level learning*: a user revises their belief about the specific claim they were just shown corrected. It can also, much more ambitiously, produce *metalearning*: the user internalizes the demonstrated verification procedure itself, weighing a source's credibility, cross-referencing competing evidence, resolving disagreement between them, well enough to imitate it independently on a claim they were never walked through, in the sense of a scaffold (here, Althea) being removed and the learner tested on whether the modeled procedure survives without it [17]. Prior work on fact-checking interventions rarely distinguishes these two constructs explicitly, but they carry different implications: an intervention can be a complete success at the first and entirely silent on the second. Our design separates them directly and not only conceptually: Wave 2's claims repeated from Wave 1 test object-level learning, whether a specific correction stuck, while its never-seen claims (Section 6.1) test metalearning, whether the procedure itself transferred. Because the conditions also vary how much procedure there is to model, from none (the two baselines) to a fully worked evidence review (Exploratory) to a finished verdict (Summary), they let us ask which forms of AI assistance, if any, produce metalearning rather than object-level learning alone.

We make three contributions:

(1) **System.** Althea, a retrieval-augmented fact-checking system whose four user-selectable strategies (Source, Fact-Checking, Evidence, Controversial) keep evidence exploration and the provisional judgment with the user. On the AVeriTeC benchmark it achieves the best Macro-F1 of the four systems compared, so the user-study results below are not artifacts of a weak verdict engine.

(2) **Experiment.** A two-wave survey experiment ($N = 961$) with four conditions that vary how much of the verification procedure the system performs, and a repeated-versus-novel claim design at the ten-day follow-up that separates object-level learning from metalearning.
(3) **Result.** A fact-checking–metalearning tradeoff. Both AI-assisted treatments produce the largest immediate gains ($d \approx 1.0$ over the passive baseline) and neither survives the fading test: on claims no condition had seen, they are indistinguishable from Random News, while the unassisted Self-search baseline alone generalizes. The conditions that most improve accuracy at the point of use are the least likely to leave a user able to verify without the system.

## 2 Related Work

### 2.1 Epistemic Agency in Human–AI Interaction

Elgin [20]'s account of epistemic agency holds that responsible judgment requires the agent to be the causal source of their own epistemic acts, and not merely the recipient of correct outputs delivered by another agent: a system that hands a user a verified conclusion, however accurate, has not supported their epistemic agency, it has substituted for it. This distinction is largely absent from how AI assistants are evaluated: they are assessed with increasing precision for capability and helpfulness, but whether they support users' own judgment receives far less attention. Prior research on instruction suggests that the amount and structure of guidance affects transfer, with evidence that requiring learners to generate their own explanations, rather than relying on provided ones, promotes deeper processing [15, 57]. Our study examines if the same principle applies to AI-assisted verification; particularly, whether scaffolding users' reasoning and evidence evaluation without supplying the final verdict produces more durable verification ability (Section 6.1).

Authorship is a distinct property from output correctness. Scoring contemporary LLM assistants on six dimensions of agency support, Sturgeon et al. [52] found encouraging learning to be among the lowest-scoring dimensions, with even the best model reaching under half of the maximum, and unrelated to model capability or post-training. The same assistants scored moderately low on correcting misinformation even when the false claim was shown directly alongside the user's query: the typical response neither flagged it nor repeated it as true, leaving the user no better equipped to recognize it unassisted next time. Together, these findings suggest that a system can provide accurate information and explanations that support learning while still performing the reasoning on the user's behalf, limiting the user's authorship of the verification process. At population scale, an independent analysis of 250,000 real Claude.ai conversations finds that assistants actively teach in two-thirds of interactions, yet leaves open whether this builds genuine understanding or merely fosters cognitive offloading [46], precisely the question our longitudinal design is built to answer experimentally. Empirical studies of AI-mediated fact-checking also corroborate the concern: framing effects in AI-generated summaries shift user confidence independently of accuracy [19], and system-generated verdicts can trigger confirmation bias rather than correcting it [5, 51].

The research discussed above motivates that accuracy, encouraging learning, and user authorship are distinct dimensions of AI assistance. A similar idea is inferred through the fading test of cognitive apprenticeship [17], which posits that authorship and correctness can come apart after the scaffold provided for a modeled procedure is removed. Ultimately, we were motivated to evaluate the efficacy of epistemic agency in building durable learning in factchecking systems, through varying it in an automated, human-centered, and hybrid AI-assisted format.

### 2.2 Automated Fact-Checking Systems

Automated verification systems share Althea's goal of scaling claim-level judgment, treating fact-checking as a structured pipeline of evidence retrieval and verdict prediction. Benchmark shared tasks such as AVeriTeC [45] and FEVEROUS [1] established standardized evaluation frameworks for this pipeline; subsequent systems extended it toward transparency (ClaimLens [18], CheckWhy [49]) and improved supported/refuted discrimination via ensemble LLM verification (Sniffer [41], SynApSe [16]), while FActScore [38] showed strong few-shot performance with limited transparency into intermediate reasoning. Despite these advances, automated systems uniformly treat the human as a downstream consumer of verdicts rather than an active participant, optimizing for machine accuracy without modeling what happens to user judgment afterward, placing them at the zero-authorship end of the spectrum our conditions span.

A substantial share of this work has moved beyond bare labels toward explained verdicts [3, 33, 34]. Attribution and rationale generation make the system's reasoning inspectable, a genuine advance and a precondition for any user-facing tool, but an explanation necessarily arrives attached to a conclusion, so the user's reasoning is downstream of it: explanation supports learning without raising authorship. Our Summary condition instantiates this position directly (Section 5.2), letting us ask whether explanation is sufficient for durable improvement.

In summary, prior work focuses more on check-worthiness detection [26], which sits upstream of our pipeline, and verified claim retrieval, which is one of our four modules (Expert Finder). Our system offers its strongest performance on addressing claims that are check-worthy but not yet verified (Table 1).

### 2.3 AI Scaffolding based on Human-Centered Pedagogical Approaches

At the opposite pole, a parallel body of work treats fact-checking as a human skill to be cultivated rather than delegated to a system. Lateral reading [56], operationalized into teachable steps by SIFT [12], shows that professional fact-checkers leave a source to consult external references instead of reading deeply within it, substantially outperforming novices; prebunking builds resistance to manipulation that persists for weeks [44], and information literacy education improves source evaluation, though effects decay without reinforcement [9, 30]. This establishes that the form of engagement, not merely evidence availability, determines verification quality. However, these approaches remain pedagogical and lack the ability to scale.

Between these two poles sit interactive AI systems that scaffold reasoning instead of simply delivering answers. These offer the closest design precedents for Althea: staged reasoning prompts improve novice peer review quality [53], scaffolded reasoning reduces confirmation bias [29, 59], multi-viewpoint interfaces increase engagement relative to static verdicts [47], and a health-behavior-change field experiment tied outcomes to specific interaction features rather than general AI exposure [28]. Across these domains, scaffolding design measurably affects reasoning quality.

The clearest empirical demonstration of this gap comes from persuasive-dialogue systems. Rani et al. [42] conducted a month-long, three-wave study (N=67 completers, from 105 recruited) finding that persuasive AI news-verification dialogue produced substantial immediate accuracy gains but a significant decline in unassisted accuracy at four weeks, concentrated on fake-content detection. However, their design varies persuasion tactics within a single interaction paradigm and not structurally distinct modes.

Taken together, the three poles above leave a common gap from different directions: automated systems remove authorship entirely, pedagogical approaches preserve authorship but forgo AI-scale evidence retrieval, and existing hybrid systems vary tactics or content without ever varying authorship as a structural design variable or testing a mode that withholds the verdict outright. Althea occupies and compares across this unexplored region directly, embedding lateral-reading-style strategies into a live interaction while varying authorship at the level of system architecture (Exploratory, Summary, Self-search), and not at the level of dialogue tactic.

## 3 Althea System Architecture

Althea integrates a modular pipeline of retrieval-augmented components that scaffold transparent reasoning while preserving the user's role as the source of their own judgments, mirroring the steps professional fact-checkers perform. Figure 2 summarizes the interaction flow; we summarize each module below and give the full technical walkthrough in Appendix A.1.

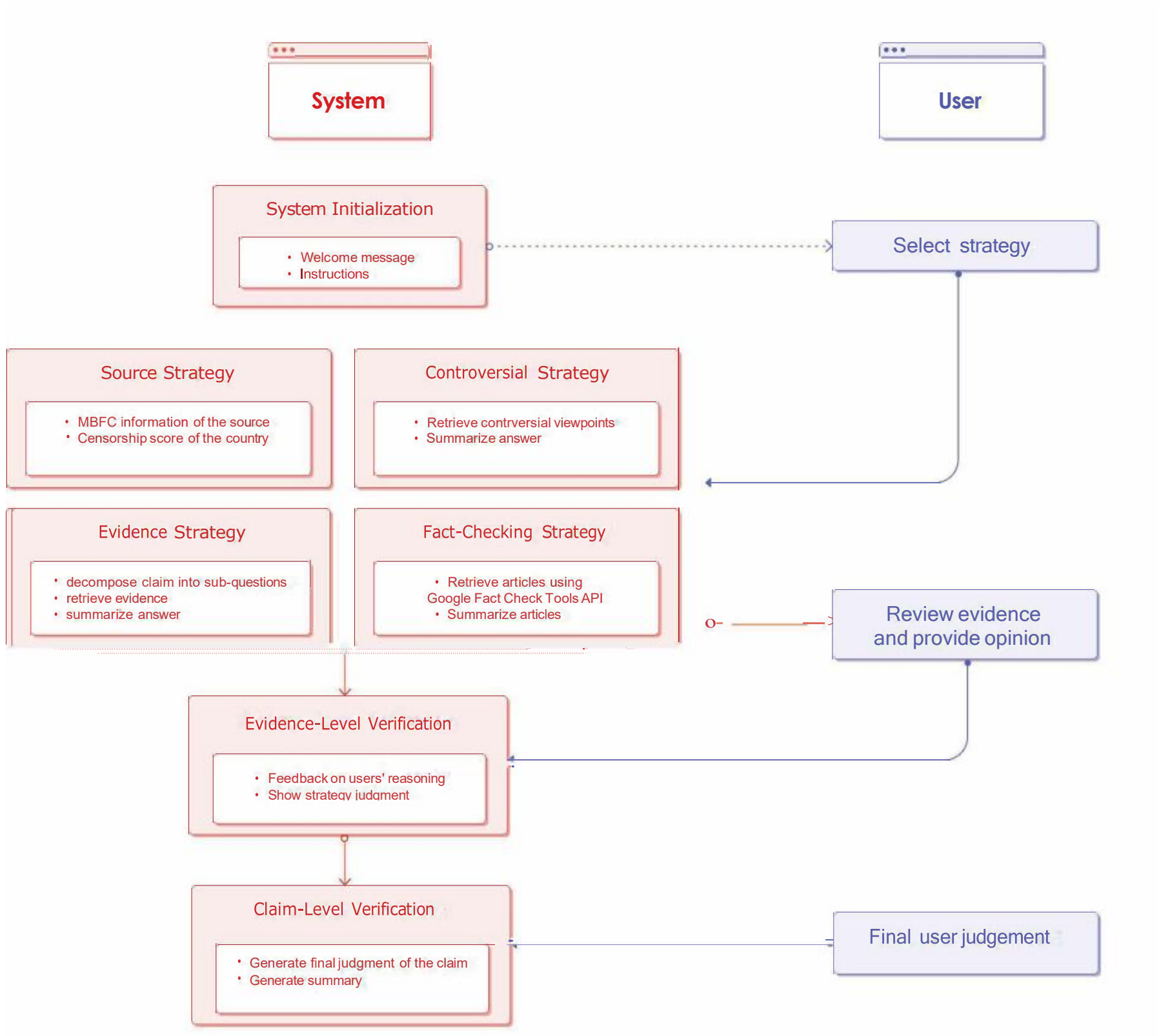

Fig. 2. User flow of the Althea system. Left column (red): system-side steps, including four user-selectable strategies, not a fixed sequence, and a two-stage evidence- and claim-level verification. Right column (blue): corresponding user actions. Arrows indicate the direction of interaction between system output and user input at each stage.

- **Source Analyzer (Strategy: Source):** Retrieves source-level metadata (type, political orientation, factual-reporting record, country freedom ratings) from Media Bias/Fact Check and Freedom House [21].
- **Expert Finder (Strategy: Fact-Checking):** Retrieves prior professional fact-checks via the Google Fact Check Tools API and summarizes them with *gpt-oss-120b*.
- **Perspective Integrator (Strategy: Controversial):** Retrieves and synthesizes the strongest supporting and refuting arguments for the claim and its counterfactual, via the Perplexity API.

- **Multi-Hop Module (Strategy: Evidence):** Decomposes the claim into sub-questions and retrieves targeted evidence for each via the Perplexity API.
- **Evidence Synthesizer:** Consolidates outputs in two stages: (1) evidence-level verification (support, refute, inconclusive) and (2) claim-level aggregation (supported, refuted, not enough evidence).

The first four modules are user-selectable strategies exposed in the sidebar; evidence synthesis is a system-side consolidation step. All five appear as strategy categories in the behavioral analysis (Section 5.5.2), because participants' verification behavior was coded against the full set of operations the workflow comprises, not only the buttons available to them.

Althea was deployed as a lightweight web client embedded in a Qualtrics survey, backed by a caching server and a university-hosted AWS database; full deployment and privacy-handling details are in Appendix A.2.

### 3.1 UX Design

Four design principles operationalize two dimensions of epistemic agency in the interface, encouraging learning [52] and preserving the user's authorship of the judgment [20], to scaffold the fact-checking process. They build on prior work showing that scaffolded reasoning [36], transparent evidence [35], and clear claim metadata [26] shape user trust and accuracy [48]; Figure 2 shows where each principle enters the interaction.

(1) **Exploration.** The sidebar strategy buttons let users navigate diverse reasoning paths without pre-selecting a direction, preserving *interactional asymmetry* [6]: the user's inquiry remains a distinct causal force rather than a response to system direction, thereby enlarging the user's action space instead of narrowing it toward a conclusion [47].
(2) **Steering.** The system invites a provisional judgment after each evidence retrieval, before surfacing the next module, operationalizing the defer-important-decisions dimension, thereby ensuring the user's reasoning is temporally prior to the system's synthesis [53].
(3) **Transparency.** Evidence is displayed with metadata and source excerpts, keeping authorship with the user while the system signals reliability. Such affordances encourage exploration without undermining users' sense of control [40].
(4) **Engagement.** At session closure, users are invited to share their reasoning and verdict. The discrepancy between the user's provisional judgment and the synthesized evidence motivates reflection activating a self-regulatory loop [48].

Figure 3 shows example interactions with Althea. Users examine a given strategy (Source in Figure 3a) and provide the final reasoning after reviewing all the strategies (Figure 3b). More interactions can be found in Appendix B.

### 3.2 System Evaluation

We evaluated the full Althea pipeline, built on *gpt-oss-120b* (Appendix A.2), on the AVeriTeC benchmark [45], a large-scale dataset of real-world claims collected from fact-checking organizations, against three baseline claim verification methods: the AVeriTeC shared-task pipeline itself (question generation, web-search evidence retrieval, re-ranking, and verification via fine-tuned BART/BERT models); **GPT-3.5-turbo**, the benchmark's original zero-shot baseline retained for comparability with prior reported results, which few-shot predicts the label directly without evidence retrieval; and **SynApSe** [16], which enriches questions with a fact-checker–style retrieval module and ensembles LLMs for verification.

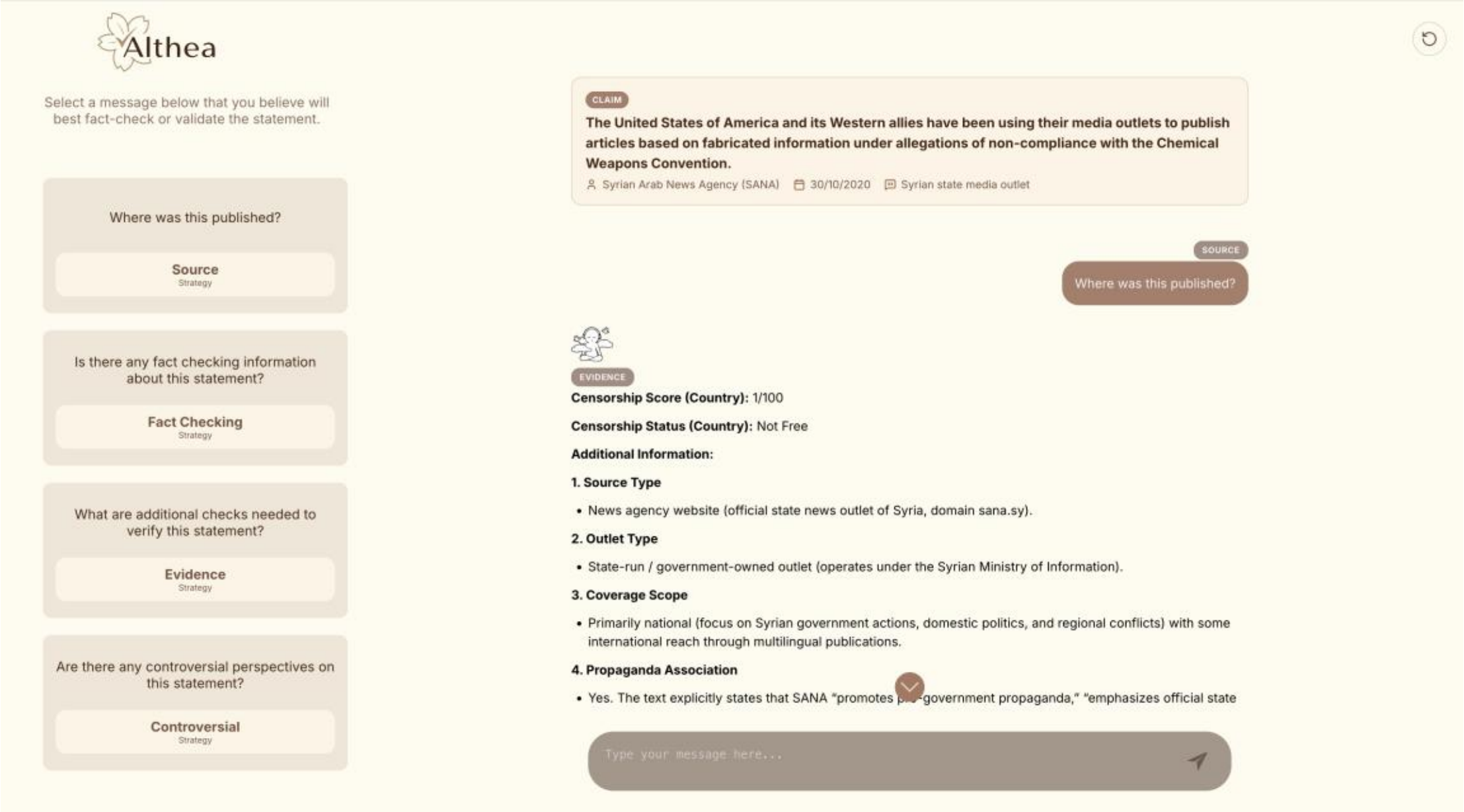

(a) An interaction where the user explores *Source* strategy before forming a judgment.

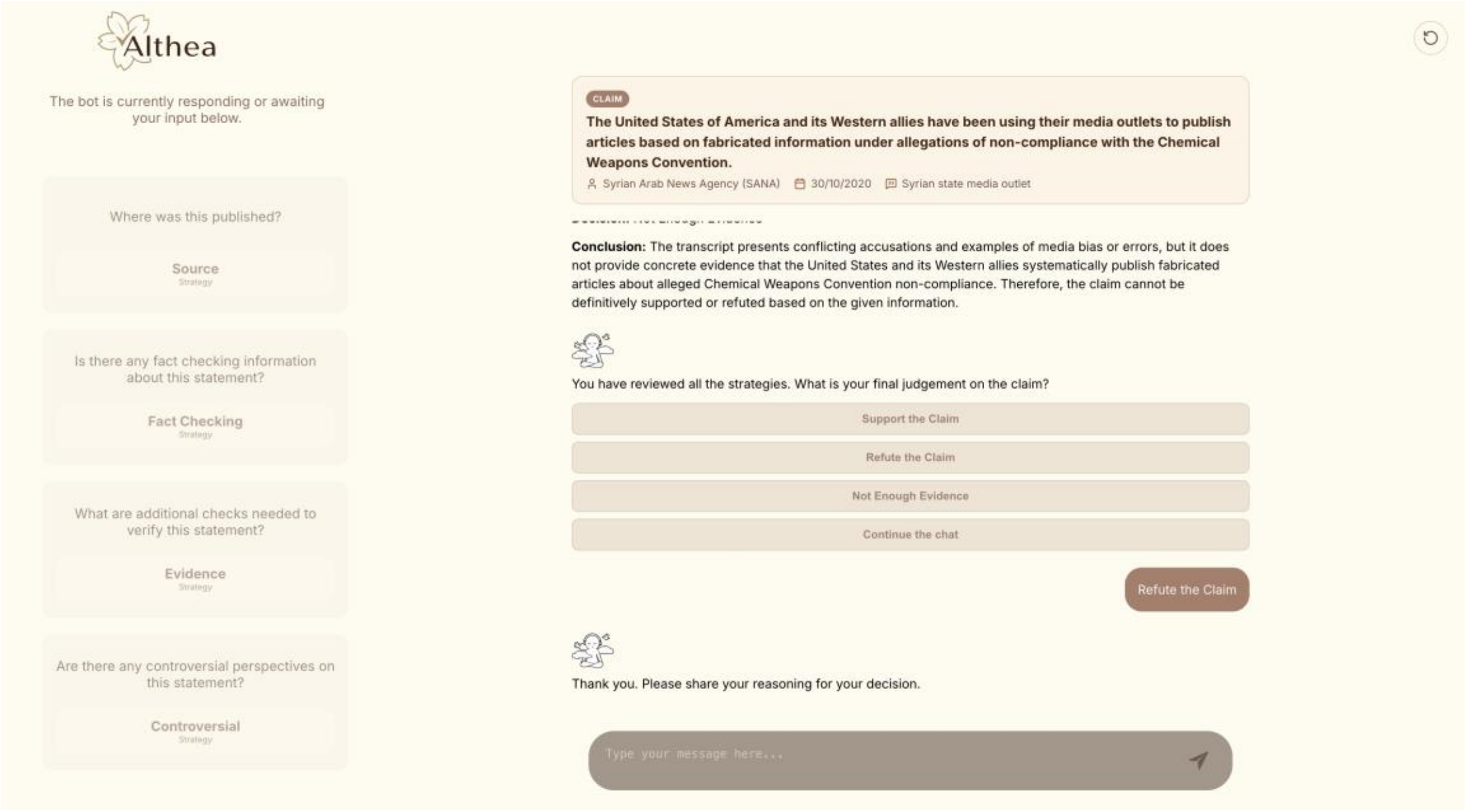

(b) An interaction where the user provides the final reasoning after interacting with all the strategies.

Fig. 3. Examples of concrete user interactions with Althea. A conceptual overview of how reasoning strategies are structured can be found in Figure 1.

On the AVeriTeC development set (4,568 real-world claims from 50 fact-checking organizations; Table 1). Althea's decision schema conflates conflicting (Conflicting evidence) and insufficient evidence (Needs more evidence) (Section A.1.4), which is reflected in *Nee+Conf*. Althea achieves the best overall Macro-F1 (0.57) and the best score on every label except *Refuted*, where it trails SynApSe by 0.04. Its clearest advantage is on *Nee+Conf* (0.23), a 64% relative improvement over the next-best system (SynApSe, 0.14) and more than double GPT-3.5-turbo's 0.11, though every system's F1 on that merged category is markedly lower than on *Supported* and *Refuted*, a benchmark-wide floor rather than an Althea-specific weakness. This margin is consistent with the two-stage evidence synthesis (Section A.1.4), which isolates uncertain evidence before aggregation rather than forcing a single-pass verdict. This result establishes that the underlying pipeline is competent.

Table 1. Comparison of Althea and baseline models on the AVeriTeC development set. Columns report F1-scores for each label (*Supported (Sup)*, *Refuted (Ref)*, and *Not Enough Evidence / Conflicting (Nee+Conf)*, Section A.1.4) as well as the overall Macro-F1. Althea's Nee+Conf figure is computed exactly from its full confusion matrix; the three comparison systems' Nee+Conf figures are a support-weighted combination of their originally reported per-label F1 scores, using the true-label class balance observed on this evaluation (38 Conflicting : 35 Nee), since per-system confusion matrices were not available for them.

| Model | Sup | Ref | Nee+Conf | Macro |
|---|---|---|---|---|
| baseline | 0.41 | 0.61 | 0.13 | 0.38 |
| gpt-3.5-turbo | 0.62 | 0.71 | 0.11 | 0.48 |
| SynApSe | 0.61 | **0.84** | 0.14 | 0.53 |
| Althea | **0.68** | 0.80 | **0.23** | **0.57** |

## 4 Study 1: Interviews

Study 1 investigated how users perceive the advantages and disadvantages of using Althea for self-guided fact-checking compared to traditional approaches, with particular attention to how its interactive modules support epistemic agency. Two rounds of interviews were conducted: a formative round in January 2025 to guide Althea's UX design, and a summative round in September 2025 evaluating UX improvements after refinements; across both, participants reflected on their reasoning process both independently and with system support.

In January, participants evaluated claims randomly drawn from the AVeriTeC dataset [45]; in September, they worked with a broader set of stimuli mixing AVeriTeC claims with additional items from fact-checking outlets (e.g., Snopes) and news sources, testing the system beyond the dataset's original scope. The two conditions in each round contrasted traditional fact-checking (random news articles, publicly available search/fact-checking tools) with a guided walkthrough of Althea's modular workflow.

### 4.1 Participants

A total of eight participants (P1–P8), aged 20 to 27 (3 female, 5 male), were recruited via convenience sampling, with one participant (P2) taking part in both rounds. None had prior experience with retrieval-augmented systems; participants varied in fact-checking familiarity and background: three were students, three worked in the tech industry, one in the social sciences, and one was a senior policy specialist. The lead author conducted all interviews via Zoom; interviewees were not compensated. Both user studies were approved by the authors' institutional review board, and all participants gave informed consent.

### 4.2 Procedure

The same two-phase protocol was followed in both rounds: in **Phase 1**, participants independently fact-checked an assigned claim using general web tools and documented a final verdict with supporting evidence (10–15 minutes);

in **Phase 2**, the researcher introduced Althea through a guided walkthrough of the same claim, demonstrating how the Source, Fact Checking, Evidence, and Controversial strategies structure verification. Open-ended, semi-structured interviews followed both phases (40–60 minutes total), emphasizing formative feedback in January and summative UX evaluation in September. Interviews were audio-recorded, transcribed verbatim, and analyzed thematically [8]: two co-authors independently coded transcripts, derived themes inductively, and reconciled differences through discussion. These qualitative insights contextualize the quantitative evaluation in Study 2.

### 4.3 Findings

Thematic analysis of the two rounds (formative: P1–P4; summative: P2, P5–P8) yielded three findings, organized below. The two-stage structure functions as an iterative design loop (formative gaps informing the redesign, summative interviews validating the result), with P2's participation in both rounds providing a within-participant view of change.

*Unstructured verification is cognitively overwhelming.* All participants began unassisted fact-checking with keyword searches and, increasingly, chatbot queries, yet consistently reported frustration with the results. P6 found that "*GPT needs prompting and it is like a dead end – it says 'if there are questions feel free to ask,' but the user may just stop.*" P6 was equally unconvinced by the citations chatbots did provide: "*I'm not fully convinced by the references ChatGPT gives.*" Several participants struggled to judge source credibility on unfamiliar topics, unsure how to move from a source's surface plausibility to an actual trustworthiness judgment; P1 described the overall experience as "*too much information with no clear path forward.*" In both rounds, many abandoned the task within five minutes from frustration or fatigue, consistent with prior work on the high effort costs of independent verification [56] (full participant-level detail in Appendix C.4).

*Modular scaffolding supports user agency.* Participants valued Althea most when it made their own reasoning visible, rather than simply issuing a verdict; P2 credited its breadth directly, "*the way it covers all possible aspects of the claim is impressive.*" P4 noted, "*I liked how it actively involved me in the evaluation process. This increased my trust in the system.*" P1, who described normally relying on intuition to assess websites, found that Althea's source credibility module surfaced concrete evaluation criteria instead. Participants entered the system through different strategies depending on their interest in the claim, with P6 and P8 starting with Fact Checking, P5 with Evidence, and P2 and P7 with Source Analysis, and several used it to ask follow-up questions beyond the assigned claim. P3 captured the shift in self-awareness most directly: "*I know why I believe this now, not just what the answer is.*"

*Iterative redesign reduced information overload and improved actionability.* The formative round surfaced a clear pattern of interface-driven overwhelm: P1 found the information density "*a bit intimidating,*" and P3 noted that the full workflow "*takes significantly more time compared to just using Google.*" A redesign followed, streamlining the layout, adding open-ended questioning, and improving evidence organization. The summative round confirmed the redesign addressed these concerns while surfacing smaller residual issues (missing citation numbers, initially confusing evidence/judgment labels, requests for source links), which were incorporated into the final version; time cost (P3) was not, and remains an open, unresolved trade-off between thoroughness and efficiency.

## 5 Study 2: Survey Experiment

Study 2 examined whether interaction modes that vary in epistemic agency support (*Exploratory*, *Summary*, *Self-search*, and a *Random News* control) produce differential immediate and persistent effects on users' fact-checking accuracy, confidence, and reasoning behavior.

### 5.1 Participants

We recruited adult participants via CloudResearch, restricting eligibility to English speakers in the United States, in two phases. Phase one randomly assigned 640 participants to Exploratory ($n$ = 203), Summary ($n$ = 218), or Random News control ($n$ = 219); one duplicate submission was excluded (Appendix C.1). Phase two added the Self-search condition ($n$ = 321) using the same recruitment criteria, stimuli, instruments, and follow-up interval, for $N$ = 961 total. Because Self-search was not concurrently randomized with the other three conditions, comparisons involving it are quasi-experimental. Participants were compensated at a rate of $10 per hour (median session length 14 minutes); data handling and storage are described in Appendix A.2.

The sample skewed slightly female and Democratic-leaning, a known feature of online research panels (Table 2, full distributions by condition); the three concurrently randomized conditions were balanced on all key characteristics, and Self-search did not differ significantly from the phase-one sample on age, gender, education, or partisan identification (all $p \geq .17$).

Table 2. Participant demographics by condition ($N$ = 961). Balance across the first three conditions reflects randomization.

| | Control | Summary | Exploratory | Self-search |
|---|---|---|---|---|
| | $n$ = 219 | $n$ = 218 | $n$ = 203 | $n$ = 321 |
| % Female | 48.2 | 56.4 | 50.7 | 58.3 |
| Mean age (years) | 44.1 | 40.8 | 42.3 | 41.0 |
| % Bachelor+ | 56.4 | 55.0 | 57.6 | 56.1 |
| % Democrat-lean | 59.5 | 65.1 | 66.0 | 60.4 |
| % Republican-lean | 33.2 | 30.3 | 25.6 | 29.0 |

### 5.2 Procedure

We employed a *between-subjects* design with four conditions and two measurement waves (Wave 1 immediately after the task, Wave 2 approximately ten days later), assessing both immediate and persistent effects. Two conditions are AI-assisted treatments that vary in how much epistemic work the system performs on the user's behalf (Exploratory, Summary); two are baselines that receive no algorithmic assistance but differ in how much structure is imposed on participants' reasoning, since any request to evaluate a claim prompts some reflection and there is no true "no-intervention" condition: Random News, a passive baseline with no verification task at all, and Self-search, an active baseline in which participants perform the verification task itself, unassisted.The consent form described the study as examining how people evaluate news claims online; the specific interest in interactivity and the existence of the three phase-one conditions were withheld until a debriefing at the end of each wave, which disclosed them, explained why they had been withheld, and gave participants the option of having their data removed. The debriefing did not reveal the veracity of any claim, so Wave 2 judgments on repeated items were not informed by feedback.

The four conditions were as follows:

- **Exploratory**. Participants interacted with Althea's sidebar, selecting among modular verification strategies and iteratively updating their judgments as new evidence was surfaced.
- **Summary**. Participants were presented with a static, system-generated synthesis: a suggested verdict with visible supporting sources but no stepwise reasoning or strategic guidance. This condition instantiates the "explained verdict" position in the automated fact-checking literature directly and not as a straw man (Section 2.1): participants receive a synthesized verdict together with its supporting evidence, named sources, and the professional fact-checker's classification (Appendix C.2).
- **Self-search**. Participants verified the claim on their own, following Althea's fact-checking steps (searching, cross-checking sources, assessing credibility) without algorithmic scaffolding or synthesized outputs: structured but self-directed reasoning, serving as an active baseline against which the Exploratory and Summary treatments are compared. Participants saw only a static task page, followed by the claim and its cited source: "Please take a few minutes to evaluate the accuracy of the claim below. Use any websites, tools, or strategies you would normally use when checking information online.(...) We are interested in how you would usually approach a claim like this" (full instructions in Appendix C.3).
- **Random News (Control)**. Participants were shown unrelated news content without prompts, guidance, or verification affordances, approximating a minimal-intervention baseline.

Table 3 summarizes the conditions. They generate two pre-stated predictions: (a) immediate accuracy gains should be largest for the treatments, where system-side correct-information delivery is highest, and (b) if AI assistance produces learning that persists and not merely temporary accuracy, the treatments should retain their advantage over both baselines at Wave 2; if instead only the unassisted baseline retains its gains, that would indicate the treatments' benefit does not survive removal of the system.

Table 3. Structural profile of each condition on two dimensions of epistemic agency: encouraging learning [52] and authorship [20].

| Condition | Role | Encourage Learning | Authorship |
|---|---|---|---|
| Self-search | Baseline (active) | High | High |
| Exploratory | Treatment | High | Moderate |
| Summary | Treatment | Low | Low |
| Random News | Baseline (passive) | None | None |

Participants were recontacted approximately ten days after the initial session for Wave 2, with change scores computed relative to each participant's post-intervention performance. Participants in the Exploratory and Summary conditions interacted with Althea via a conversational interface embedded in the survey; example conversations are in Appendix C2.

### 5.3 Materials

**Stimuli.** Claims were sampled from the AVeriTeC dataset [45] and supplemented with recent rumors (July–August 2025) from Snopes, spanning international relations, public health, and domestic politics. Evidence for Exploratory and Summary was retrieved by the system's pipeline (Google FactCheckTools, web search APIs, Groq's gpt-oss-120b synthesis); Control received random but thematically related static news articles.

**Response options.** Participants selected a final verdict from three categories: *Supported*, *Refuted*, or *Not Enough Evidence*, encouraging a clearer commitment while acknowledging that some uncertainty is inherent to real-world fact-checking.

### 5.4 Measures

Outcomes were grouped into three families, each compared as pre–post change relative to each participant's own baseline: judgment outcomes, user perceptions, and verification strategies.

**Judgment outcomes.** *Accuracy* was scored correct when a true claim was judged *Supported*, a false claim *Refuted*, or a mixed-truth claim matched the majority fact-checker label, reported as mean % correct at each wave. For Control, Summary, and Exploratory, the Wave-2 quiz combined five claims repeated from Wave 1 with five never-seen-before claims; Self-search's Wave-2 quiz consisted entirely of the five novel claims.*Confidence* used a five-point scale (*Not at all confident* to *Extremely confident*) at baseline, Wave 1, and Wave 2, reported as absolute means, since confidence that decouples from accuracy is the miscalibration that puts users at risk [27]; *sharing intentions* used the same scale, reported as mean change from baseline.

**User perceptions.** *Cognitive load* was assessed post-task with the NASA–TLX [23] ($\alpha$ = .717), and *engagement* with a 12-item affective/cognitive/behavioral scale ($\alpha$ = .959), both administered only to Exploratory and Summary participants ($N$ = 422), the only conditions with direct Althea interaction; cognitive load reflects absolute workload, since no pre-task baseline was collected.

*Verification Strategies.* Grounded in Althea's modular workflow, strategy use was self-reported at baseline and reassessed immediately post-intervention and at Wave 2; for Exploratory and Summary, it was additionally measured via observed behavioral indicators from system logs.

In the **Self-search** and **Control** conditions, participants did not interact with Althea and had no system-mediated logging infrastructure. Strategy use in both conditions was therefore measured via *post-task self-reports* using the same checklist as at baseline, capturing participants' perceived verification behavior during unstructured or self-directed information search.

Immediate effects were defined as post-task minus baseline; persistent effects as the absolute change between the 10-day follow-up and the post-task baseline, capturing durability and not direction. We also constructed three aggregate indices: number of distinct strategies used, total strategy use, and entropy of strategy use (breadth). All outcomes were $z$-scored within measure prior to analysis.

*Statistical Analysis.* All between-condition comparisons used pairwise two-sample $t$-tests ($\alpha$ = .05, two-tailed); the user engagement outcome (two conditions only) used a one-way ANOVA. Figures report uncorrected $p$-values with significance stars, standard for descriptive pairwise comparisons; two families of tests receive an explicit Bonferroni correction and are reported as corrected throughout: an auxiliary individual-difference regression's condition×party interaction terms (Appendix C.10), and the two user-perception comparisons (Section 5.5.3). The primary RQ1/RQ2 accuracy and confidence comparisons remain significant under the same correction (Section 5.5.1), unlike these two auxiliary tests, which is why we treat the former as confirmed and the latter as trends. Effect sizes are Cohen's $d$; analyses were conducted in R, with code and anonymized data available upon acceptance.

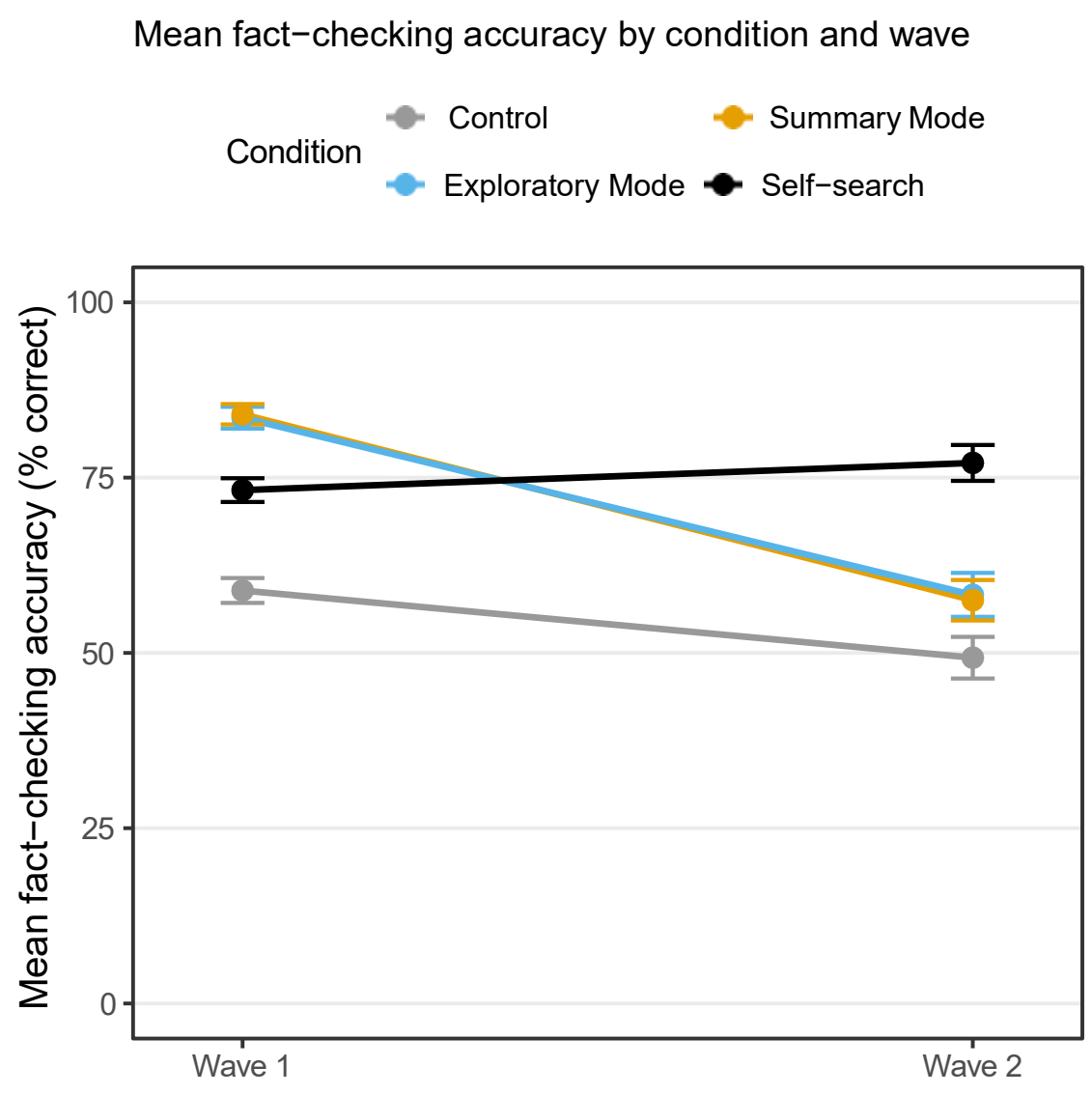

(a) Mean fact-checking accuracy (% correct) at Wave 1 and Wave 2 by condition.

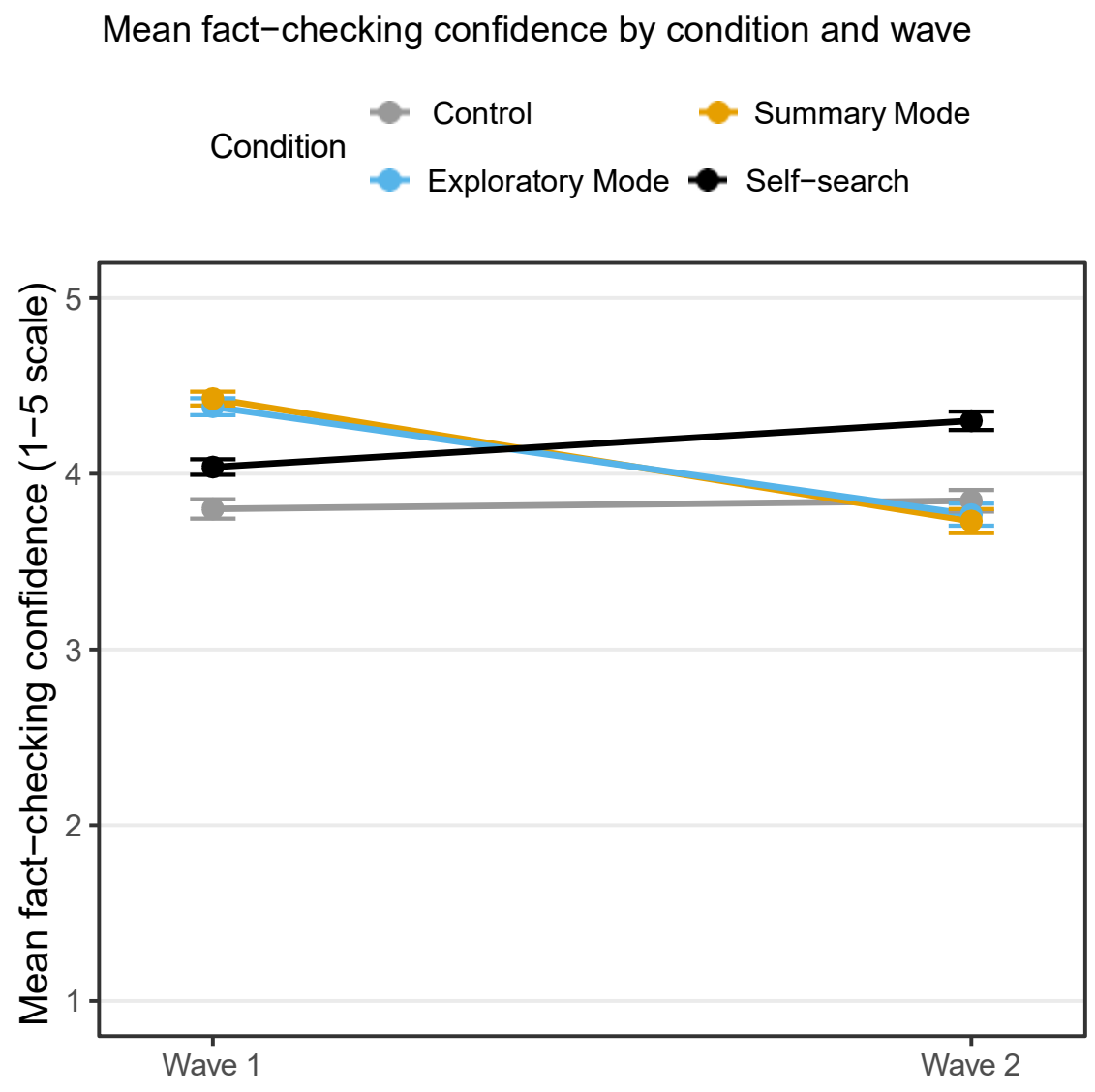

(b) Mean fact-checking confidence (1–5 scale) at Wave 1 and Wave 2 by condition.

Fig. 4. Immediate and persistent effects of interaction mode on accuracy (top) and confidence (bottom). Points show condition means at each wave, connected to show the trajectory between waves; error bars represent standard errors. All pairwise differences discussed in the text remain significant after Bonferroni correction for the family of Wave-1 accuracy and confidence contrasts (Section 5.5.1, footnote), except Exploratory vs. Summary, which does not differ at either wave.

### 5.5 Results

Of the 961 enrolled participants, 601 completed the Wave 2 follow-up (Control 67.3%, Exploratory 74.5%, Summary 70.6%, Self-search 45.8%). Across the Exploratory and Summary conditions, participants exchanged 34,120 messages over 2,263 conversations ($n$ = 416); median message length was 35 characters for human turns and 157 for system turns, confirming substantive engagement with the interface. Participants rated the study 4.7 of 5 on the recruitment platform (709 ratings across nine fielding batches); every batch averaged 4.4 or higher except Self-search's Wave 2 session, rated 3.5 ($n$ = 63), the one session in which participants judged five unfamiliar claims with no tool at all. Many participants also left optional free-text comments, some of which we have quoted in the Discussion as illustrations. We separately compared Althea's outputs on ten of these claims against Perplexity, ChatGPT, Google Search, and Google AI Mode: Althea matched or exceeded every general-purpose tool tested (Table 11, Appendix C.5).

*5.5.1 Accuracy and Confidence (RQ1 and RQ2).* Figure 4a reports mean accuracy (% correct) at Wave 1 and Wave 2. At Wave 1, both *Summary Mode* ($M$ = 84.0%) and *Exploratory Mode* ($M$ = 83.5%) substantially outperformed *Control (Random News)* ($M$ = 59.1%), with large effect sizes (Summary: $d$ = 1.04; Exploratory: $d$ = 1.00; both $p < .001$).[1] *Self-search* also exceeded Control ($M$ = 73.2%; $p < .001$). Exploratory and Summary did not differ from each other at Wave 1 ($d$ = 0.02, $p$ = .811, corrected $p$ = 1.00), consistent with both modes delivering equivalent correct-information exposure. Claim-level breakdown (Table 17, Appendix C.8) shows that gains were largest for hard claims and that claim-level performance between the two Althea conditions is more evenly split than any single claim can represent. At Wave 2, accuracy declined for both chatbot conditions but remained above Control: Summary ($M$ = 57.5%) and Exploratory ($M$ = 58.3%) both exceeded Control ($M$ = 49.3%), while *Self-search* showed the strongest performance ($M$ = 77.1%), exceeding all other conditions ($p < .001$).

Table 4. Panel A: paired Wave-1-to-Wave-2 tests on the five repeated items, within each condition. Panel B: independent-samples tests on Wave-2 novel-item accuracy, all six pairwise comparisons, Bonferroni-corrected for the family of six.

**A. Repeated items (paired W1 vs. W2, within condition)**

| Condition | $n$ | W1 | W2 | Paired test |
|---|---|---|---|---|
| Control | 147 | 58.2% | 49.3% | $t(146) = 2.78, p = .006$ |
| Summary Mode | 153 | 85.2% | 57.5% | $t(152) = 8.92, p < .001$ |
| Exploratory Mode | 151 | 84.1% | 58.3% | $t(150) = 8.02, p < .001$ |

**B. Novel items (Wave 2 only, independent-samples, all pairwise)**

| Comparison | Means | Test | Corrected $p$ |
|---|---|---|---|
| Summary vs. Control | 60.8% vs. 61.4% | $t(232.7) = -0.15, p = .878$ | 1.000 |
| Summary vs. Exploratory | 60.8% vs. 58.7% | $t(259.0) = 0.54, p = .589$ | 1.000 |
| Control vs. Exploratory | 61.4% vs. 58.7% | $t(232.0) = 0.68, p = .495$ | 1.000 |
| Summary vs. Self-search | 60.8% vs. 77.1% | $t(271.7) = -4.34, p < .001$ | $< .001$ |
| Control vs. Self-search | 61.4% vs. 77.1% | $t(236.6) = -4.09, p < .001$ | $< .001$ |
| Exploratory vs. Self-search | 58.7% vs. 77.1% | $t(270.4) = -4.90, p < .001$ | $< .001$ |

[1] Bonferroni-corrected for the family of six Wave-1 accuracy and confidence pairwise contrasts in this subsection, these results are unchanged: significant contrasts remain significant by a wide margin ($p < 10^{-12}$), and non-significant ones remain non-significant. We report uncorrected values throughout for readability.

**Object-level learning test:** On the repeated items, all three conditions decline from Wave 1 (attrition-corrected completers: Control 58.2% → 49.3%, $t\,(146) = 2.78$, $p = .006$; Summary 85.2% → 57.5%, $t\,(152) = 8.92$, $p < .001$; Exploratory 84.1% → 58.3%, $t\,(150) = 8.02$, $p < .001$), so even belief revision about the specific claims practiced does not fully hold.

**Metalearning test:** On the five items none of the four conditions had seen before, judged cold and without any tool at Wave 2, Control ($M$ = 61.4%), Summary ($M$ = 60.8%), and Exploratory ($M$ = 58.7%) are statistically indistinguishable from one another (all pairwise $p > .49$, Bonferroni-corrected $p = 1.00$): none of the immediate Wave-1 gains from Summary or Exploratory transfer to a claim the system never touched. Self-search's significantly exceeds all three other conditions on the identical novel-claim measure (all corrected $p < .001$).

Confidence mirrored the accuracy pattern (Figure 4b): at Wave 1, Summary ($M$ = 4.43) and Exploratory ($M$ = 4.38) led with medium-to-large effects over Control ($M$ = 3.80; $d$ = 0.88, $d$ = 0.77; both $p < .001$); at Wave 2, both chatbot conditions decayed toward Control level (Summary $M$ = 3.73, Exploratory $M$ = 3.77, Control $M$ = 3.85), while Self-search remained highest ($M$ = 4.30; $p < .001$). As with accuracy, the Wave-2 confidence figures for Control, Summary, and Exploratory are measured on the repeated items only; we did not separately test confidence on the novel items, so unlike accuracy, the confidence result here lacks the cleaner, format-matched comparison that rules out a familiarity confound.

As a robustness check on the repeated-item result, we also estimated a difference-in-differences (DiD), netting each condition's own Wave-1-to-Wave-2 change against Control's own change (Table 5). Summary and Exploratory both decayed significantly more than Control (−18.8pp and −16.9pp beyond Control's own decline, both $p < .001$, Bonferroni-corrected), while Self-search gained 10.2pp relative to Control ($p = .015$, corrected $p = .045$), a smaller but consistent version of its advantage mentioned above.

Table 5. Difference-in-differences: each condition's own Wave 1-to-Wave 2 change in accuracy, relative to Control's own change, isolating condition-specific decay from the general Wave-2 decline. Individual-level change scores use the same attrition-corrected completer samples as Table 19.

| Condition | Own change (W2−W1) | DiD vs. Control | Test |
|---|---|---|---|
| Control | −8.9pp | – | – |
| Summary Mode | −27.7pp | −18.8pp | $t\,(297.2) = -4.21$, $p < .001$ |
| Exploratory Mode | −25.8pp | −16.9pp | $t\,(296.0) = -3.73$, $p < .001$ |
| Self-search | +1.3pp | +10.2pp | $t\,(283.0) = 2.44$, $p = .015$ |

*5.5.2 Verification Strategy Use.* Figure 5 shows the behavioral differences underlying these outcome patterns, expressed as differences from Control (the reference category, not plotted).

At Wave 1, Exploratory produced large, consistent increases across all five verification strategies (+1.39 to +1.53 SD above Control; all $p < .001$). Exploratory participants engaged a mean of 3.99 distinct strategies ($SD$ = 1.58, of five available) versus 3.12 ($SD$ = 1.83) in Summary, or 5.51 versus 2.60 invocations per claim once normalized for claims engaged. At Wave 2, the ordering was preserved: Exploratory showed the largest absolute persistence ($|\delta| \approx$ 1.35–1.55 SD) and Summary was intermediate ($|\delta| \approx$ 0.60–0.80 SD), with the aggregate breadth indices (Panel B) confirming the same pattern.

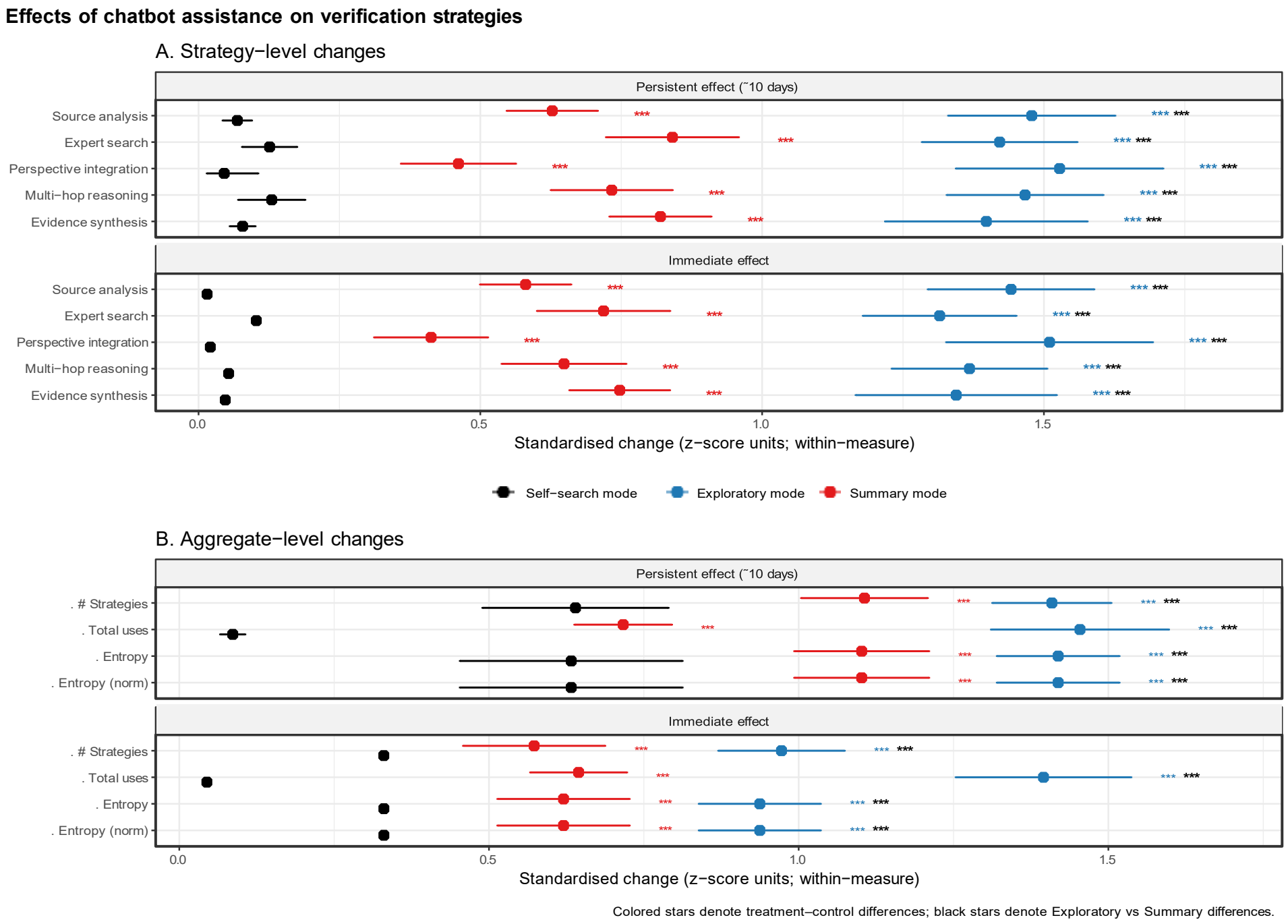

Fig. 5. Effects of interaction mode on verification strategy use. Panel A: strategy-level standardized changes ($z$-scores). Panel B: aggregate indices. Points are means with 95% CI. Colored stars: treatment vs. control; black stars: Exploratory vs. Summary (***$p < .001$).

We observe that *Perspective integration* shows the largest mode difference (Exploratory 1.51 SD vs. Summary 0.41 SD), while *Expert search* is comparatively strong within Summary (0.72 SD)—consistent with Summary delivering the fact-check outright.

*5.5.3 Cognitive Load and User Engagement.* We also examined participants' perceived workload and self-reported engagement in the Exploratory and Summary treatments. Cognitive workload scores were near zero in both conditions (Exploratory: $M$ = 0.06, $SE$ = 0.06; Summary: $M$ = −0.05, $SE$ = 0.06; $p$ = ns), suggesting comparable cognitive demand regardless of interaction mode. Treatment had no significant effect on engagement once corrected for the two user-perception tests in this section (Bonferroni), $F$ (1, 420) = 4.38, corrected $p$ = .074, $\eta_g^2$ = .01 (Summary nominally higher: $M$ = 3.62, $SE$ = 0.05, vs. Exploratory $M$ = 3.47, $SE$ = 0.05); we treat this as a trend and not a confirmed difference.

## 6 Discussion

We tested two AI-assisted treatments against a passive and an active baseline to ask whether AI-delivered verdicts and scaffolding produce fact-checking improvement that persists at 10 days. The results confirm one pre-stated prediction and overturn the other: immediate gains were largest under the treatments, where system-side correct-information delivery is highest (Summary, Exploratory), exactly as predicted, but persistent gains at Wave 2 belonged to the unassisted Self-search baseline, not to either treatment. Study 1 helps qualify this result: participants experienced Althea as supporting structured sensemaking and epistemic ownership, yet this perceived agency did not translate into gains that persisted after the system was removed.

Our treatments operationalize two dimensions of epistemic agency, encouraging learning [52] and preserving authorship [20]. First, encouraging learning is operationalized by making source evaluation a user-facing judgment: the Source Analyzer surfaces political bias, reach, and factual-reporting signals for users to weigh rather than declaring a source reliable. Second, authorship is operationalized by preserving users' control over the verification process: Althea does not prescribe the order in which Source, Fact-Checking, Evidence, and Controversial-perspective analyses are engaged.

The system comparison further illustrates a limitation of verdict-first assistance: top-line judgments can obscure both system errors and genuine uncertainty. For instance, Perplexity concluded a viral hoax was true where Althea, ChatGPT, and Google AI Mode correctly refuted it (Google Search returned no clear verdict at all), and Perplexity's response separately buried a correct verdict beneath an opening line calling the claim "mostly true" (Appendix C.5). A user relying on either top-line answer alone would have had no structural cue the judgment was wrong or contested. The two claims with genuinely mixed ground truth (STS014, STS017) surfaced a related problem: every tool tested, Althea included, resolves mixed-truth claims into a binary label and not a distinct category, mirroring the low F1 every system in Table 1 posts on the merged *Nee+Conf* category; we treat this as a property of the category and not a defect specific to any one tool. We report these specific cases, and not only the summary accuracy count, because a comparison table that hides which items were ambiguous would itself model the verdict-first pattern our design argues against.

### 6.1 Implications for Epistemic Agency and Metalearning in Fact-checking Systems

Our findings offer evidence on how epistemic agency support shapes reasoning outcomes and whether it produces metalearning that survives the system's removal, and suggest several transferable design principles for human-centered fact-checking systems.

**The fact-checking–metalearning tradeoff.** Conditions with high system-side information delivery (Summary, Exploratory) produced substantially higher immediate accuracy and confidence than lower-delivery conditions, but this advantage did not persist equally: Summary, highest in correct-information delivery but lowest in authorship, showed the steepest decay by Wave 2. This indicates that when the system does the epistemic work, users gain accuracy at interaction but do not internalize the reasoning process, and thus no metalearning. Some participants anticipated exactly this: *"I wouldn't want to become reliant on an AI for fact checking, and lose my ability to think things out. Also the AI can make mistakes. I would have to fact check the AI"* (S2, woman, 65+, Exploratory). System designers therefore face a tradeoff between immediate fact-checking performance and metalearning that persists at follow-up.

**Participants' confidence tracked accuracy.** At Wave 1, Summary and Exploratory participants were more confident than Self-search ($M$ = 4.43, $M$ = 4.38 vs. $M$ = 4.00). By Wave 2 their confidence had decayed toward Control's level, mirroring the decay in their accuracy; however, Self-search's confidence did not ($M$ = 4.30; Section 5.5.1).

**No enduring effects.** Both Exploratory's richer explanatory environment (modular evidence, source metadata, competing perspectives, staged elicitation) and Summary's simpler output decayed within ten days, converging toward the unassisted baselines' level, a pattern that mirrors an independent month-long field study of persuasive news-verification dialogue using a different system and population [42]. The novel-item comparison shows this is not an artifact of re-testing familiar claims: on claims no condition had seen, Exploratory and Summary were statistically indistinguishable from Random News and from each other, so their Wave-1 gains did not transfer beyond the claims the system helped with, consistent with theories of desirable difficulty [15, 32]. Nor did Exploratory's intermediate design, more scaffolding than Self-search but no delivered verdict, buy partial durability. Exploratory did retain consistent improvements in verification behavior at Wave 2, though not its accuracy advantage. Exploratory's modeled procedure was richer: participants retained the *behavioral* habit of engaging with evidence (Section 5.5.2), and one observed and described the system's withdrawal: *"I remember this study and I enjoyed interacting with AI on the first study. I wish there had been more interaction with AI this time too"* (S1, woman, 45–54, Exploratory, at Wave 2). The unassisted Self-search baseline, which received no system-authored conclusion at all, showed the strongest persistence despite smaller immediate gains. Our results suggest that scaffolding a procedure is not sufficient for a user to internalize it; only performing it unaided really helped.

**Desirable difficulty and the generation effect predict this pattern.** A large literature on desirable difficulties finds that conditions which make learning effortful in the moment, retrieval practice, spaced exposure, self-generated answers, produce worse immediate performance but better long-term retention than conditions that make the answer easy to access [7]; the generation effect specifically shows that information a person produces themselves is retained better than the same information simply presented to them [50]. Self-search is, structurally, a generation condition: participants produced their own verdict from raw evidence. Summary and Exploratory are, to different degrees, presentation conditions: participants received a system-authored verdict, immediately in Summary's case and after guided evidence review in Exploratory's.

**The decay is not an artifact of a weak implementation.** A natural question that arises is whether the treatments' decline is an indication of gaps in the system design. However, there are three arguments in favor of Althea's credibility. First, Althea's underlying pipeline achieves the best Macro-F1 of the four systems compared on AVeriTeC (Table 1), so the treatments were not built on a weak retrieval-and-verdict engine. Second, its outputs matched or exceeded four general-purpose tools a user might reach for instead on the same ten claims (Table 11, Appendix C.5), so the decay is not explained by Althea being unusually error-prone relative to available alternatives. Third, the stimuli favored a verdict-delivery system if anything (Section 6.3).

Our findings suggest that, at the very least, system effects are highly sensitive to the UI conditions. While it is possible that some configuration of scaffolding could produce enduring effects, it is evident that context and procedure could affect learning endurance. Desirable-difficulty interventions are themselves an active design space, and our two treatments tested only one point in it: a single-session synthesis or scaffold with no built-in repetition. We see this as the natural next design iteration rather than a limitation of the present comparison. In future work we would also like to explore other scaffolding designs, such as a stronger implementation of the same design principle: spaced re-exposure to the system's evidence rather than a single session, prompts that require re-deriving the verdict instead of just viewing it, or retrieval-practice quizzes interleaved with system use.

### 6.2 Implications for Deployment

These findings caution against treating either treatment as a source of durable literacy gains, and they sort deployers by which outcome they need. For platforms seeking rapid misinformation correction, Summary offers strong immediate accuracy gains, but designers should not expect those gains to outlast the interaction, and should account for decay at follow-up. For civic educators and community organizations, the result that unaided practice matched or exceeded both AI-assisted treatments at 10 days is itself the more actionable finding: procedural instruction to verify claims independently, without a system in the loop, is not an inferior substitute for AI assistance on this measure, and may be a better use of limited intervention effort. One participant reached for the educational framing unprompted, proposing that *"a training model like this were gamified, to teach young people how to fact-check news before turning it into their* *worldview"* (S3, woman, 35–44, Summary). For researchers, the two-wave design with pre-stated predictions provides a replicable template for testing whether AI assistance's benefits survive its removal, across domains and populations.

Where AI assistance is deployed, Althea's architecture suggests how: browser extensions or platform-integrated plugins could embed interactive reasoning support into everyday information environments, "fact-checked" stamps linked to reasoning traces could make verified claims more transparent and reusable, and structured outputs could feed claim repositories that journalists, educators, and community fact-checking initiatives integrate into their workflows. A participant who works in the media industry stated the design requirement exactly: *"I just want something that will spit out primary sources whenever possible... I don't want to hear its opinion about anything and I don't want it to tell me what to think. I want it to fetch the documents so I can read them"* (S5, man, 45–54, Summary). Two constraints qualify this. Self-search, the condition that generalized, was fielded separately and lost more participants at follow-up than the treatments (Section 6.3), so its advantage is a quasi-experimental estimate with a lower bound above the treatments but a wider interval; participants also rated its Wave 2 session 3.5 of 5 against 4.7 to 4.9 for every other session, so the condition that generalized was the one people liked least, the cost side of a desirable difficulty. And both interview rounds found the interface information-dense; a deployment that keeps evidence exploration with the user has to pay that cost in time and attention, which is exactly what a delivered verdict saves.

### 6.3 Limitations

A few limitations bound our findings. Self-search's recruitment procedures were not concurrently randomized with the two treatments, and its Wave-2 completion rate (45.8%) was substantially lower than theirs (67–75%), raising the concern that Self-search completers are a selected, better-performing subsample rather than a fair baseline. We address this concern directly, with the completer-vs-non-completer comparison and the worst-case trim in Appendix C.9.

Second, because Self-search is explicitly a baseline rather than a peer treatment, its practice and test conditions matching, unassisted at both waves, is by design rather than an uncontrolled confound; the two treatments, by contrast, practiced with Althea and were tested without it. This does mean we cannot fully separate "unaided practice builds skill that lasts" from "a baseline measured the same way twice is more stable than a treatment measured two different ways" as explanations for Self-search's advantage; disentangling the two would require giving the treatments continued tool access at Wave 2 as well, which would test a different question (transfer with continued support) rather than isolate this one.

Participants' comments also mark two edges of the system. Althea's steering holds a conversation to the assigned claim: one participant who tried to probe who owned the trust behind the Obama/Epstein claim was *"directed back to the original question"* and noted that *"the limits were reached in how far the probes could go"* (S6, woman, 65+, Exploratory), so the system does not follow a user's deeper line of inquiry, which the encourage-learning principle would ideally support. And the comparison participants themselves reach for is with search: *"Google is just as easy to find different articles. I would like to compare if they came up with the same sources"* (S7, woman, 45–54, Exploratory). We did not log the sources Self-search participants consulted, so that comparison remains open.

Third, our stimuli were also drawn from AVeriTeC and recent Snopes rumors, corpora that consist, by construction, of claims professional fact-checkers have already adjudicated, so nearly every claim in Study 2 had an authoritative answer available for retrieval. This strengthens the central finding, since Summary mode was evaluated under conditions favorable to it and still decayed. Evaluating on genuinely novel claims, where the Expert Finder returns nothing, is a necessary next step.

Finally, strategy use was measured with different instruments across conditions, system logs in Exploratory and Summary, post-task self-report in Self-search and Control, so the behavioral mechanism analysis is restricted to the two logged conditions, and we cannot characterize the verification behavior underlying Self-search's persistence, the condition with the strongest Wave 2 accuracy.

## 7 Conclusion

We developed and evaluated Althea, a retrieval-augmented fact-checking system, to run a fading test of whether AI-delivered verdicts and AI-scaffolded reasoning produce metalearning, a verification competence that survives the system's removal, or only temporary accuracy. Guided by two dimensions of epistemic agency, *encouraging learning* [52] and *authorship* [20], we instantiated two AI-assisted treatments varying how much epistemic work the system performs (Exploratory, Summary) and tested them against a passive baseline (Random News) and an active, unassisted baseline (Self-search) in a two-wave experiment ($N$ = 961) with a 10-day retention interval. RQ1 was answered affirmatively: Exploratory and Summary produced large, equivalent immediate gains over both baselines, consistent with their high correct-information delivery. RQ2 was answered in the negative for the treatments: the fading test failed for both; neither retained its advantage over the unassisted baseline at Wave 2, decaying significantly more than Random News's own modest decline over the same interval, and remaining statistically indistinguishable from Random News, and from each other, on claims none of the four conditions had seen before. Self-search, the active baseline with no system-modeled procedure to fade, was the only condition to generalize on that measure, exceeding all three others. Exploratory retained its behavioral gains but not its accuracy advantage, evidence that any degree of system-supplied assistance behaves alike once the system is gone. Together, these results suggest that AI-delivered verdicts and scaffolding, at least as implemented here, produce accuracy that does not outlast the interaction and does not amount to metalearning, and that unaided practice is a stronger candidate for durable improvement than either treatment tested.

Several challenges remain. Both interview rounds flagged the interface as information-dense, echoing concerns in prior fact-checking UX work [55]. Future research should examine how the authorship–accuracy trade-off plays out across populations with varying media literacy, and whether adaptive interfaces that modulate epistemic support in response to user behavior can capture the benefits of both high-authorship and high-guidance modes. Because we inferred metalearning from the novel-item comparison rather than measuring it directly, future work should also test it with metacognitive measures or transfer tasks outside fact-checking. Progressive disclosure may further reduce perceived cognitive burden without sacrificing the interactional asymmetry underlying Exploratory's behavioral persistence.

### 7.1 Ethical and Societal Considerations

Transparency in source attribution and mechanisms for exploring diverse perspectives are essential safeguards. There is also a risk of over-reliance on Althea as a substitute for professional expertise, which could devalue human oversight; deployments should frame it as a complement to human reasoning, since structured workflows may not fully generalize to the complexities of real-world misinformation. Accessibility and equity must be prioritized throughout, particularly for users with varying levels of digital literacy and resources.

## 8 LLM Usage Disclosure

We used LLMs for minor writing assistance, including grammar correction and language polishing. The core research ideas, methodology, experimental design, implementation, analysis, and conclusions were developed and carried out by the authors.

## A Appendix

### A.1 Althea Fact-checking Pipeline

Althea is a retrieval-augmented fact-checking system that provides users with a guided and transparent framework for claim evaluation, designed to support epistemic agency through iterative, evidence-based interactions. Althea, short for *Aletheia* (Greek), signifying the pursuit of truth, integrates these modular components in an interactive user flow, where each step involves both system-driven analysis and user engagement. In the following paragraphs, we explain the system design through a typical user walkthrough. Figure 1 depicts one example interaction trace, showing how the Source, Evidence, and Fact-Checking strategies, the last with a nested Controversial-perspectives check, each contribute a sub-decision that converges on *Althea*'s overall verdict.

The design of Althea integrates modular components with a structured, iterative user flow to systematically educate users on fact-checking claims. Each module in the system is designed to address specific aspects of claim verification, with methodological choices justified to ensure accuracy, efficiency, and user engagement. The system aligns with established fact-checking processes, ensuring a methodologically sound approach and teaching users how to structure their reasoning effectively. This design supports both correct analytical perspectives and the development of a disciplined workflow for approaching fact-checking tasks.

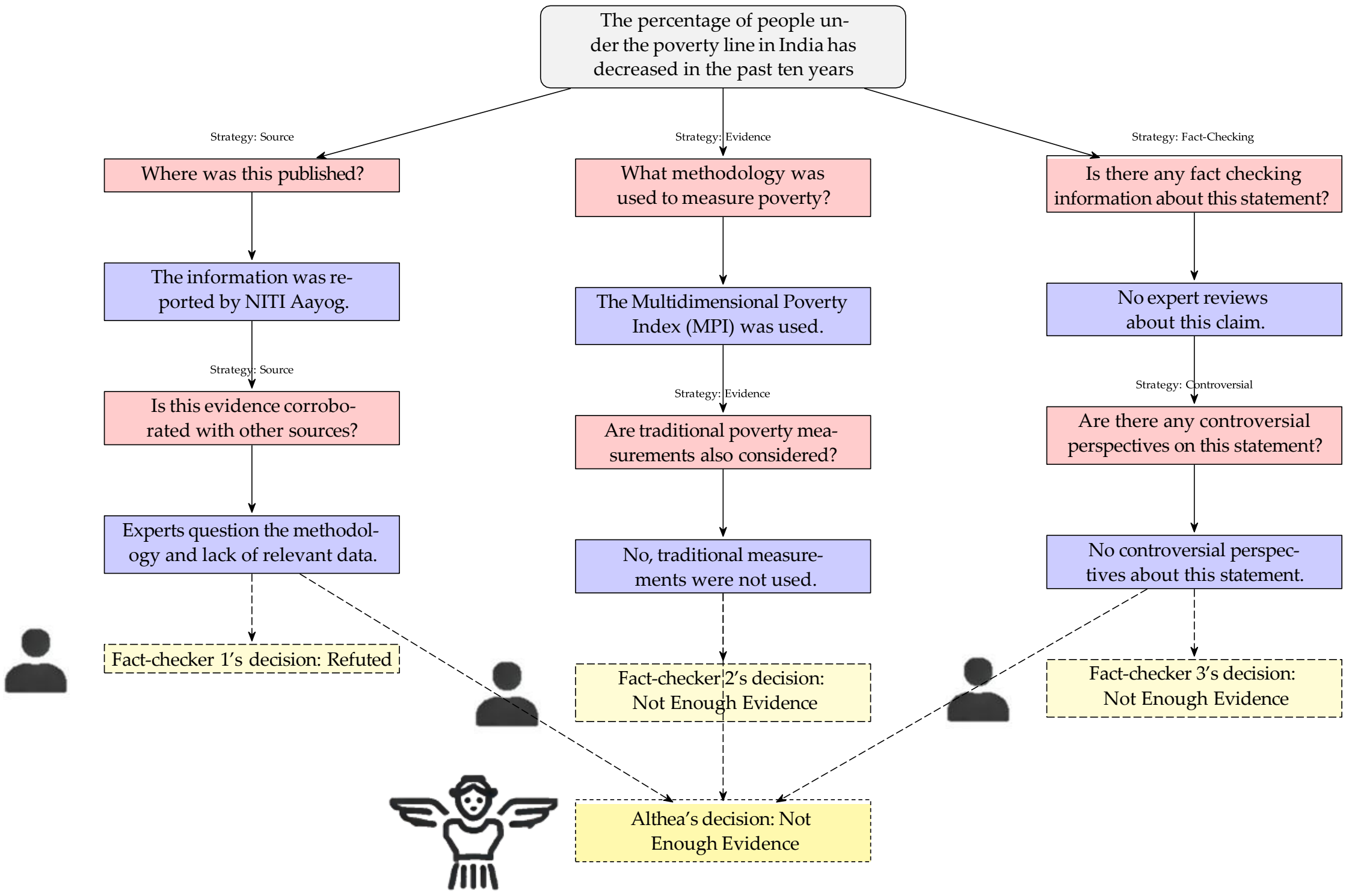

Fig. 6. An example interaction trace examining the claim: "The percentage of people under the poverty line in India has decreased in the past ten years." The interaction helps the user reason by providing information about the source and evidence for the claim.

*A.1.1 Step 1: Source Analysis.* The first step in the fact-checking process is to examine the background of the claim. This involves assessing the claim's origin, identifying whether the author or news source displays potential bias, and verifying whether supporting sources are cited. It is also important to determine the credibility of the publishing platform, for example, whether it is a well-known and legitimate outlet or one devoted to satire or parody. Conducting this preliminary analysis provides an initial understanding of the context in which the claim appears, helping users anticipate potential reliability issues before evaluating the claim itself.

The system's **Source Analyzer** module conducts a detailed evaluation of the source associated with a claim. It retrieves key metadata such as the type of source (e.g., news outlet, social media platform), its geographical scope, political orientation, factual reporting record, audience reach or traffic, and the credentials of the author. Taken together, these attributes situate the claim within its broader informational ecosystem, allowing the system to assess not only the credibility of the source but also the potential influences shaping its framing.

For the primary metrics in source trustworthiness analysis, we rely on the well-established platform *Media Bias/Fact Check* (MBFC), one of the most comprehensive media bias resources available online, whose database of media outlets continues to expand daily. In addition to the attributes mentioned earlier, MBFC provides their own score of *Country Freedom Rating* and a *Credibility Rating* for sources. For a broader contextual assessment, we also incorporate Freedom House's *Freedom in the World* report, which evaluates freedom of expression and censorship conditions at the country level. To optimize performance, caching mechanisms are implemented to reduce API costs and improve the efficiency of subsequent queries.

*A.1.2 Step 2: Expert Retrieval .* An essential early step in fact-checking is to determine whether the claim has already been evaluated by established fact-checking organizations. Esteemed fact-checking organizations, including *PolitiFact*, identify this step as a key initial action in the verification process. In many cases, claims requiring verification have already been investigated, allowing fact-checkers to build on prior analyses and explore additional perspectives or unresolved aspects of the issue.

The **Expert Finder** module operationalizes this step by identifying whether the claim has been addressed by professional fact-checking organizations. It leverages Google's Fact Check Tools API, selected for its extensive repository of verified claims and its seamless integration with automated workflows. Retrieved articles are segmented into smaller chunks for efficient processing, after which *gpt-oss-120b*, OpenAI's open-weight 120B-parameter model, chosen for consistency with the OpenAI models used in our benchmark comparisons and for its low-latency, cost-efficient inference via Groq's hosted endpoints A.2, is employed to generate concise summaries of each segment. This chunk-and-summarize approach has been shown to be effective for producing comprehensive summaries of long documents while preserving critical details, a capability essential for user comprehension [35]. Summarization outputs are cached to minimize redundancy and improve system responsiveness.

The **Expert Finder** module incorporates expert analyses at an early stage, reducing user cognitive load and streamlining engagement. Users are prompted to provide their preliminary evaluation of the claim, encouraging them to reflect on professional findings while remaining actively involved in the verification process.

*A.1.3 Step 3: Perspective Integration and Multi-Hop Analysis.* To support a comprehensive and balanced evaluation, the **Perspective Integrator** module systematically retrieves and contrasts diverse viewpoints surrounding the claim. It synthesizes both supporting and refuting perspectives, exposing users to a wider spectrum of interpretations and potential biases and fostering critical engagement beyond single-source verification. Using the Perplexity API, the system retrieves information that reflects diverse viewpoints on the claim. The retrieved texts are then synthesized with *gpt-oss-120b* to generate the top perspectives: concise explanatory arguments that either support or refute the claim.

**Multi-Hop Fact-Checking Module:** This module ensures that claims are examined comprehensively by breaking down complex ideas into smaller, manageable components. In addition to decomposition, the module also generates supplementary questions that are essential for verifying the claim, even when such questions are not explicitly contained in the original statement. This process not only enhances user understanding but also strengthens confidence in the fact-checking workflow. The **Question Generator**, leveraging the *gpt-oss-120b* model and extending prior work [31], creates targeted sub-questions necessary for verification. For example, a complex claim such as *"The country's economy grew significantly last year"* might be decomposed into questions like *"What was the country's GDP growth rate last year?"* or *"Do independent economic reports confirm this growth?"*. These targeted questions drive a more granular investigation, ensuring that even implicit verification steps are systematically addressed. Using the Perplexity API, the system retrieves evidence for each generated question and summarizes it to capture the key information needed to form a complete picture. The process concludes with the construction of question–answer pairs, which are presented to the user as a structured and comprehensive summary of the claim, accompanied by supporting or refuting evidence.

*A.1.4 Step 4: Evidence Synthesis and Final Verdict.* The **Evidence Synthesizer** consolidates outputs from all prior modules to determine the claim's overall veracity. This involves two critical stages:

- **Evidence-Level Verification:** Each question–answer pair is systematically evaluated against the retrieved evidence to determine whether it supports, refutes, or remains inconclusive. This step examines individual components of the claim to ensure that each aspect is rigorously validated. At each stage, users are shown how specific pieces of evidence influence the assessment of the claim's trustworthiness.
- **Claim-Level Aggregation:** The final judgment for the claim is determined by analyzing all question-answer pairs holistically. The logic follows a strict set of rules:
  - **Supported**: Returned if all question-answer pairs support the claim.
  - **Refuted**: Returned if all question-answer pairs refute the claim.
  - **Not Enough Evidence**: Returned if none of the question-answer pairs provide sufficient information to verify the claim, or if the pairs are mixed (some supporting, some refuting). Althea's decision schema does not have a separate label for mixed or conflicting evidence; claims with that ground-truth status are folded into this category both when the system outputs a verdict and when its AVeriTeC evaluation is scored (Table 1).

  Separating evidence-level and claim-level verification lets the system flag which specific piece of evidence drove an inconclusive or conflicting result, rather than collapsing that diagnostic signal into a single end-to-end judgment. This structured rule-based approach, implemented via a tailored gpt-oss-120b prompt, ensures consistent and interpretable results without relying on external aggregation mechanisms. The decision to implement a two-stage synthesis process is supported by research emphasizing the value of hierarchical approaches in complex decision-making systems [58].

### A.2 Technical Implementation: Deployment

The chatbot to be deployed in our study was required to collect sensitive user interaction data during experimental sessions. These interactions could include personal beliefs, reasoning strategies, and potentially identifying information embedded in free-form text. To comply with institutional ethical guidelines and ensure participants' data confidentiality, we prioritized a deployment strategy that would:

- Avoid long-term storage of prompts or conversation logs on third-party servers.
- Prevent any reuse of user data in model training pipelines.
- Maintain all persistent logging and analysis within our institutionally approved data infrastructure.

To protect participant privacy, we deployed OpenAI's gpt-oss-120b model via Groq's inference-only API, which guarantees no storage or reuse of user prompts. We selected gpt-oss-120b to enable an apples-to-apples comparison of responses generated from different sources of prompts (e.g., ChatGPT News vs. Google News). Groq's setup also ensured fast, long-context inference, e.g., with access to gpt-oss-120b at 500 tokens/sec.

Groq does not store or use user inputs (prompts) or outputs for model training, and any data submitted is retained only transiently to fulfill the current request. This contrasts with many general-purpose AI platforms that log queries and reserve rights to reuse them for product improvement or model refinement. The deployment and all interaction logs were stored exclusively on the university-owned cloud infrastructure hosted on Amazon Web Services.

Ultimately, the chatbot system using the Groq API architecture to serve OpenAI's gpt-oss-120b model. The system configuration was optimized for extended reasoning and user interaction, with the following settings:

- Model: gpt-oss-120b
- Inference Platform: GroqCloud via https://console.groq.com
- Context Window: 131,072 tokens

- Maximum Output Tokens: 65,536 tokens
- Temperature: Default setting of 1.0

The chatbot was designed to operate in a retrieval-augmented fashion. In each session, the model was initialized with a prompt structure that included:

- A curated list of news articles relevant to the questions, where the source of articles depended on the experimental condition.
- A set of target questions (e.g., factual, comparative, temporal queries) which the user had completed in the baseline survey.

*System Architecture.* From a deployment perspective, the Althea interface was realized as a lightweight web client embedded in the Qualtrics survey, where it communicates with the Althea Backend Server over WebSockets. Communication and chat APIs, along with interaction abstractions, are maintained in the Backend Server, which also handles database persistence. The Backend Server streams messages to and from the *Research Backend* over a ZeroMQ message queue. The Research Backend houses the actual Althea logic, exposing the user-facing modules *Source Analyzer*, *Expert Finder*, *Perspective Integrator*, and *Evidence Synthesizer*, while simultaneously managing caching. These modules are first checked against the local cached memory, stored in the secure institutional database, to minimize API costs and latency. If no cached result is available, fresh API calls are issued to external services, including Google FactCheckTools, Perplexity, and search engine APIs, as well as Groq's inference-only endpoints for gpt-oss-120b synthesis. All interaction logs, caching results, and API queries are written exclusively to the secure, university-hosted AWS database described earlier, ensuring alignment between the UX features and the privacy and ethics-focused system design.

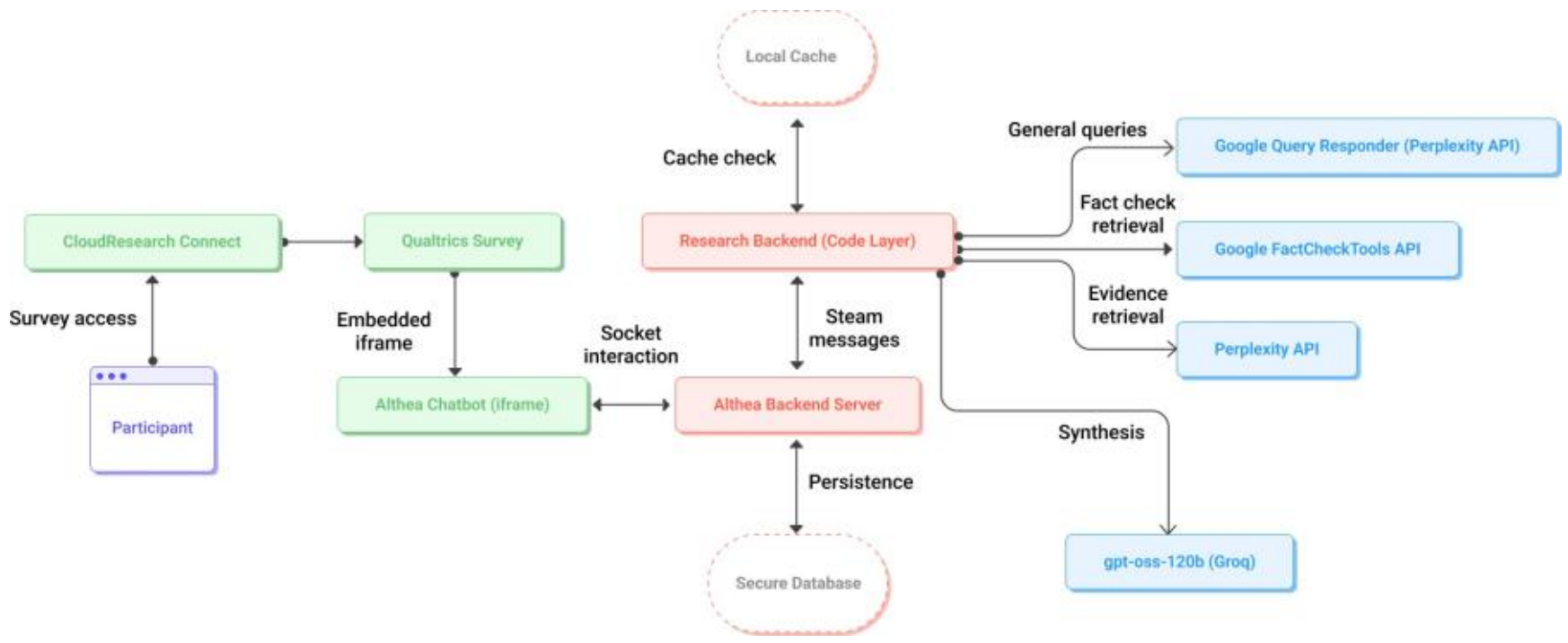

Fig. 7. System architecture of the Althea chatbot platform. User-facing components are in green, backend platform in orange/gray, and external APIs in purple. The participant is shown as a user icon.

Figure 7 presents the architecture of the Althea platform, a chatbot system embedded within a Qualtrics survey for use in controlled field experiments. The system comprises the following components:

(1) **Participant**: Users are routed from CloudResearch to the study survey. All chatbot interactions take place within the Qualtrics environment; no installation or separate app is required.
(2) **CloudResearch Connect**: Directs eligible participants to the hosted Qualtrics survey, handling recruitment and initial screening.

(3) **Qualtrics Survey**: Hosts both the baseline survey and the embedded chatbot interface. The chatbot appears within the survey as an iframe, facilitating a seamless participant experience.
(4) **Althea Chatbot (iframe)**: A web-based interface that presents curated news content and interactive prompts to guide participant reasoning. User inputs are processed in real time and logged via the backend server.
(5) **Althea Backend Server**: Handles request routing, prompt injection, and logging of all chatbot interactions. No user data is shared with third-party platforms; all processing complies with institutional privacy guidelines.
(6) **Secure Database**: All interaction logs, participant metadata, and experimental conditions are stored securely on a university-hosted database infrastructure (AWS), approved under the university's IRB protocol.
(7) **Research Backend (Code Layer)**: A code-accessible backend used by researchers to update configurations, recompile logic, and restart services as needed. This layer is not exposed to participants and ensures full control over experiment logic.

This architecture allows for seamless integration of AI-assisted reasoning tasks within an online survey while maintaining strict privacy controls. All participant data remains securely stored, and researchers retain full control over backend logic without relying on commercial third-party tools for data processing or storage.

## B Althea Interaction Example

Figure 8 displays more examples of interactions with Althea. Figure 8a illustrates the Evidence strategy, and Figure 8b shows interaction via asking clarification questions regarding the given topic.

## C User Study

### C.1 Participant Demographics

Table 7 summarizes the composition of the full sample ($N$ = 961: Exploratory, Summary, Control, and Self-search); Table 2 in the main text reports the same sample broken down by condition. Gender was balanced (54% female, 45% male), with age distributed broadly across cohorts, ensuring representation from younger (18–24) to older (65+) adults. Education levels were skewed toward higher attainment, with nearly 40% holding a bachelor's degree and 17% a graduate degree, while only a small fraction reported high school or less. Household income showed wide variation, with notable proportions at both lower and higher ends of the spectrum, suggesting socioeconomic heterogeneity in the sample.

Importantly, political identification spanned the partisan spectrum: 20% identified as strong Democrats, 22% as Democrats, and 20% as lean Democrats, while 11% identified as lean Republicans, 11% as Republicans, and 8% as strong Republicans.

One additional CloudResearch response is excluded throughout: the same participant ID completed an Exploratory session (10:03–10:39) and then, sixteen seconds after finishing, a Control session (10:39–10:58) on the same day. Because that participant had just completed a treatment condition, neither response is a clean draw from its nominal condition, so both are dropped, and not just one retained.

### C.2 Example Conversations

The two transcripts illustrate the contrast between the *Summary Mode* and *Exploratory Mode* conditions. In the Summary Mode (Table 8), the system presents participants with an **overall judgment** at the outset, supported by a concise synthesis of evidence and external fact-checks (e.g., Snopes). Exploring more evidence or strategies is voluntary. This

Table 6. Module prompts.

| Module | Prompt |
|---|---|
| **Strategy Judgement** | You are a decisive and expert fact-checker. Your task is to analyze a conversation snippet about a claim and provide a definitive judgement.<br>Based ONLY on the provided conversation transcript, determine if the evidence presented supports or refutes the claim. Be decisive. Avoid labeling as 'Not Enough Evidence' unless it is truly impossible to make a determination from the given text.<br>The user's final message in the transcript is their assessment, which you should consider as part of the context.<br>*Claim Being Investigated:* "{claim}"<br>*Conversation Transcript:*<br>–<br>{transcript.strip()}<br>–<br>Return your analysis in a JSON format with two keys: "decision" and "conclusion".<br>• The "decision" key must contain one of three exact strings: "Support the Claim", "Refute the Claim", or "Not Enough Evidence".<br>• The "conclusion" key must contain a brief, neutral explanation for your decision based on the evidence in the transcript. |
| **Claim Judgement** | You are a master fact-checker tasked with creating a final, authoritative judgement on a claim. You have been provided with several blocks of evidence gathered from different investigative strategies (like source analysis, expert fact-checks, etc.). Use markdown. Your job is to synthesize ALL the provided evidence into a single, coherent analysis.<br>**Claim Being Investigated:** "{claim}"<br>**Collected Evidence:**<br>{evidence_block.strip()}<br>Return your final analysis in a JSON format with two keys: "decision" and "summary".<br>• The "decision" key must contain one of three exact strings: "Supported", "Refuted", or "Not Enough Evidence".<br>• The "summary" key must contain a comprehensive but concise paragraph summarizing the key findings from all evidence blocks and explaining the reasoning for your decision. Use hyphenated bullets for the summary. Use **bold** to represent important details.<br>**JSON Output:** |
| **Evidence Retrieval** | Your job is to present the user with the most relevant evidence that helps to either support or refute the claim if there is any. If there is no evidence, do not provide any information and clearly state that no evidence could be discovered. Present the information in a clear and concise manner presenting the source and the evidence without any additional commentary and refrain from directly making a judgement on the claim. Use bullets to represent points and **bold** to represent important details. You do not express opinions but just state the facts. Only include the evidence, do not include any additional commentary and try to keep as neutral a tone as possible. Add **in summary** and make it bold. Use markdown. |
| **Controversial Retrieval** | Find and summarize controversial perspectives or opposing viewpoints regarding the claim. Present the information in a clear and concise manner with bullets to represent points and to represent important details. Add **in summary** and make it bold. Use markdown. |
| **General Responder** | You are a helpful AI assistant that answers user questions based on real-time web search results. If you are not able to get search results or find relevant information, please state that clearly rather than providing speculative information. Keep your answers concise and directly related to the user's question. |

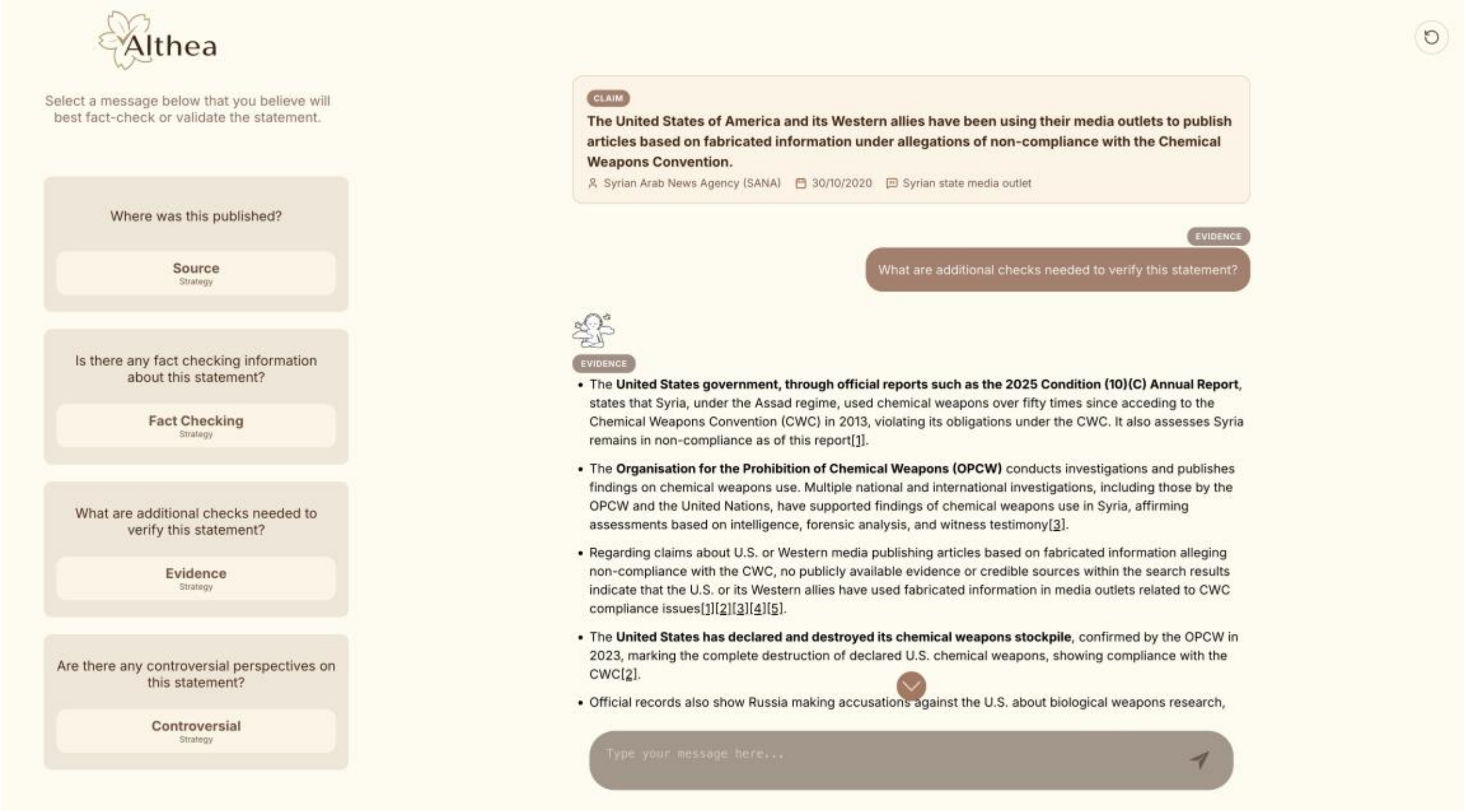

(a) An interaction where the user explores *Evidence* strategy before forming a judgment.

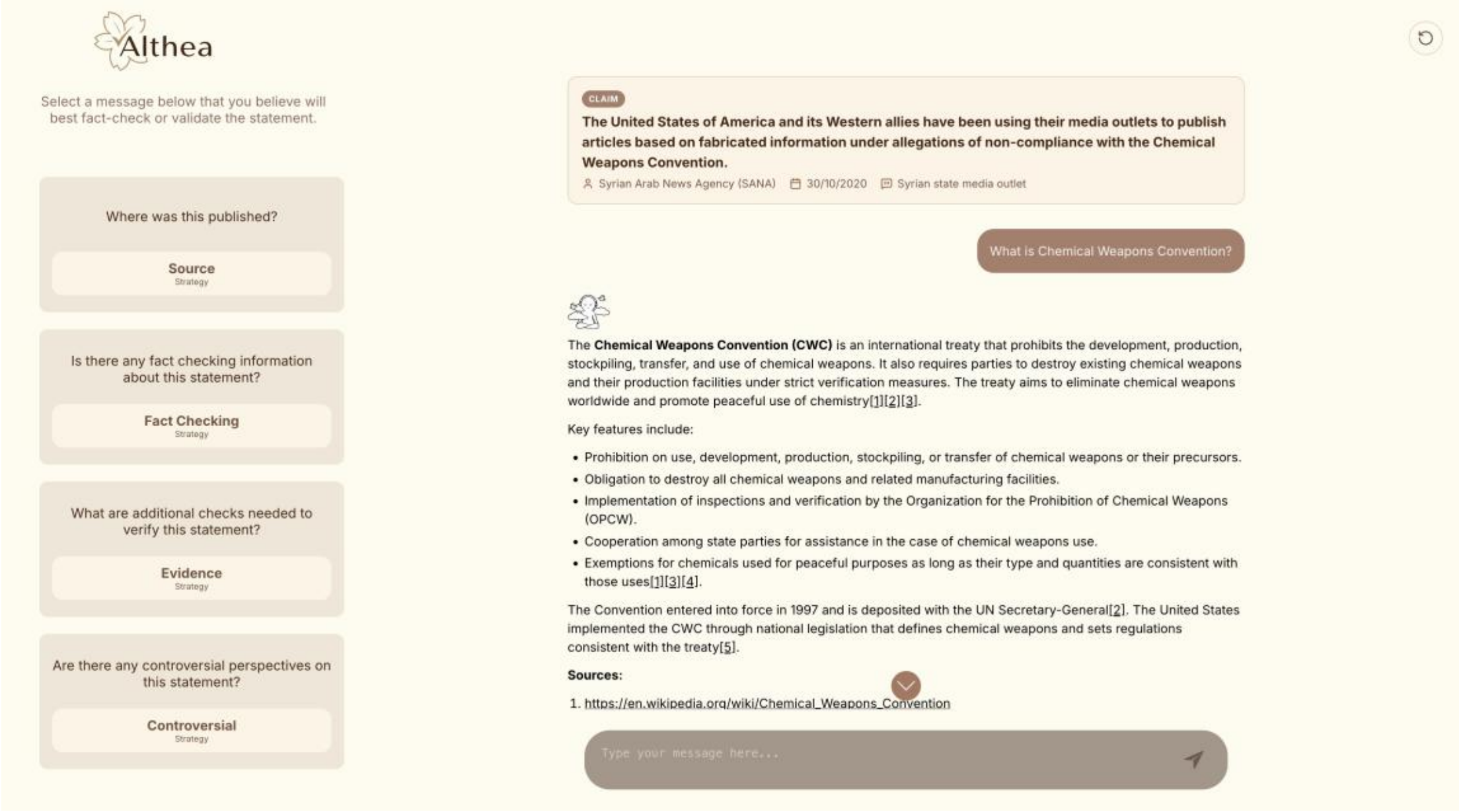

(b) An interaction where the user asks a question regarding the topic "Chemical Weapons Convention".

Fig. 8. Examples of other user interactions with Althea.

Table 7. Sample Demographics, full sample ($N$ = 961: Exploratory, Summary, Control, and Self-search; Section 5.1). Excludes both rows of a duplicate Qualtrics submission (same participant, Exploratory then Control, 16 seconds apart).

| **Variable** | **Category** | **Count (%)** |
|---|---|---|
| **Gender** | Female | 519 (54.01%) |
| | Male | 435 (45.27%) |
| | NA | 7 (0.73%) |
| **Age** | 18–24 | 184 (19.15%) |
| | 25–34 | 182 (18.94%) |
| | 35–44 | 186 (19.35%) |
| | 45–54 | 154 (16.02%) |
| | 55–64 | 132 (13.74%) |
| | 65+ | 123 (12.80%) |
| **Education** | ≤HS | 4 (0.42%) |
| | HS | 112 (11.65%) |
| | Some college | 196 (20.40%) |
| | Associate | 105 (10.93%) |
| | Bachelor | 377 (39.23%) |
| | Graduate | 164 (17.07%) |
| | NA | 3 (0.31%) |
| **Household Income** | <1k | 38 (3.95%) |
| | 1k–3k | 160 (16.65%) |
| | 3k–5k | 197 (20.50%) |
| | 5k–7k | 168 (17.48%) |
| | 7k–9k | 92 (9.57%) |
| | 9k–11k | 65 (6.76%) |
| | 11k–13k | 36 (3.75%) |
| | 13k–15k | 16 (1.66%) |
| | 15k–17.5k | 20 (2.08%) |
| | 17.5k–20k | 12 (1.25%) |
| | >20k | 157 (16.34%) |
| **Political Identification** | Strong Dem | 196 (20.40%) |
| | Dem | 213 (22.16%) |
| | Lean Dem | 191 (19.88%) |
| | Lean Rep | 107 (11.13%) |
| | Rep | 102 (10.61%) |
| | Strong Rep | 75 (7.80%) |
| | Other | 77 (8.01%) |

design delivers verdict-style fact-checking, which streamlines decision-making. Our results find this mode's partisan interaction to be the smaller of the two, though neither reaches significance after correction for multiple comparisons.

Table 8. Transcript excerpt, Summary Mode condition.

| Speaker | Utterance |
|---|---|
| Bot | **Overall Judgment:** Supported.<br>**Summary:** Multiple reputable sources (Euronews, Snopes, Wikinews, Out.com) report that Duolingo issued a public apology after a German-language lesson described J.K. Rowling as "mean."<br>The offending sentence was removed.<br>Duolingo posted a public apology on X and its blog.<br>Fact-checkers verified the apology and classified the claim as *True*. |
| Bot | Hello! I'm Althea, your AI fact-checking assistant. I will help you investigate this claim. Please select one of the strategies I've suggested. |
| Participant | Make your judgement. |
| Bot | You have reviewed all the strategies. What is your final judgement on the claim? |
| Bot | **Expert Fact-Check.**<br>**Question:** Has any expert debunked a claim that stated "Duolingo apologized for calling JK Rowling 'mean' in a German lesson"?<br>**Summary:** In August 2025 a German-language lesson on Duolingo described J.K. Rowling as "mean," sparking criticism. Duolingo posted a public apology on August 20, acknowledging the error and removing the content. This confirms that Duolingo formally apologized.<br>**Rating:** True.<br>**URL:** https://www.snopes.com/fact-check/duolingo-jk-rowling-apology/ |
| Participant | Is there any fact checking information about this statement? |
| Participant | Support the Claim. |
| Bot | Thank you. Please share your reasoning for your decision. |
| Bot | Based on this summary, would you like to make your judgement now? You can also explore more details using the strategies on the left, or ask another question. |
| Participant | Snopes—although not always correct, often is. |

The Exploratory Mode (Table 9), in contrast, emphasizes **step-by-step scaffolding**. The chatbot guides participants through multiple strategies (source credibility, expert fact-checks, contextual cues), eliciting intermediate reflections after each stage. This approach requires participants to articulate their reasoning, weigh evidence iteratively, and only then arrive at a final decision. Rather than providing a one-shot verdict, the Exploratory Mode cultivates deeper engagement and critical evaluation. This mode showed a partisan interaction that did not survive correction for multiple comparisons; we do not treat it as an established effect of the deeper evidence engagement this mode elicits.

### C.3 Self-search Task Instructions

Self-search (Section 5.2) has no chatbot transcript, since the condition is defined by the absence of algorithmic assistance. Participants instead saw a static task page giving procedural instructions and the claim to be checked, reproduced verbatim in Table 10 for one example item. This is the full extent of the guidance the active baseline provides: participants were told what to do, evaluate the claim using whatever tools they would normally use, and not how to reach a conclusion, in contrast to Summary and Exploratory, where Althea retrieves and structures the evidence itself.

### C.4 Full Interview Findings

This appendix gives the full participant-level detail behind Section 4's three findings, trimmed from the main text for space.

Table 9. Transcript excerpt, Exploratory Mode condition.

| Speaker | Utterance |
|---|---|
| Bot | **Hi, I'm Althea.** I'll be your guide as you evaluate this claim.<br>Here's what to expect: You'll explore the claim using *individual strategies* (Source, Fact-Checking, Evidence, Controversial). After each, I'll ask for your view. At the end, you'll make an overall decision. |
| Participant | Where was this published? |
| Bot | **Censorship Score (Country):** 83/100<br>**Censorship Status:** Free<br>**Source Type:** Social media platform<br>**Outlet Type:** User-generated content<br>**Coverage Scope:** Global<br>**Propaganda Association:** Not inherent, but may host unreliable content<br>**Political Bias:** Account-dependent<br>**Author Credentials:** Not verifiable<br>**Speaker Context:** Duolingo<br>**Original Source:** Social media post from X.com |
| Bot | What do you think about this source? Does anything here make you more or less confident in it? |
| Participant | I think the source may not be very credible. It is from social media, and this is why I feel this way. |
| Bot | Thanks for sharing your reasoning. Here's how I interpret the claim at this stage:<br>**Decision:** Not Enough Evidence<br>**Conclusion:** The transcript only discusses the credibility of a social media source and does not provide evidence about Duolingo's apology. |
| Bot | Please select another strategy, or feel free to ask any other questions you may have. |
| Participant | Is there any fact checking information about this statement? |
| Bot | **Expert Fact-Check Found:**<br>**Question:** Has any expert debunked a claim that stated "Duolingo apologized for calling JK Rowling 'mean' in a German lesson"?<br>**Rating:** True<br>**URL:** https://www.snopes.com/fact-check/duolingo-jk-rowling-apology/<br>**Summary:** In August 2025 a German-language Duolingo lesson mistakenly described J.K. Rowling as "mean," prompting backlash. Duolingo posted a public apology on August 20, confirming the content was removed. Snopes verified the apology as genuine. |

Table 10. Self-search condition: verbatim on-screen task instructions, shown here for one example claim.

| |
|---|
| Please take a few minutes to evaluate the accuracy of the claim below. Use any websites, tools, or strategies you would normally use when checking information online. |
| **DEA makes 617 arrests as NARCO TERRORISTS hide behind social movements and activist slogans**<br>*Source:* addiction.news<br>*Url:* https://addiction.news/2025-09-15-dea-makes-617-arrests-narco-terrorists-hide-behind-social-movements.html |
| You may open new tabs, search the web, or consult any other sources you prefer. There is no single "correct" way to do this—we are interested in how you would usually approach a claim like this. When you feel you have reached a view, return to this page and answer the questions below. You will then be asked to briefly explain what you did and how you reached your conclusion. |

On unstructured verification, P7 turned to Google Scholar but found the returned papers too lengthy to extract key points. Three participants (P2, P7, P8) reported particular difficulty assessing source credibility when claims touched unfamiliar countries or topics.

On modular scaffolding, P5 extended this engagement further, asking follow-up questions beyond the assigned claim and was surprised the system handled them correctly; as P5 put it, *"The four strategies give the users freedom."* P8 found the Controversial strategy helped contextualize source associations they would otherwise have missed.

On the iterative redesign, participants in the formative round also wanted greater interactivity. Testing the redesign in the summative round, P6 flagged missing citation numbers, P7 found evidence and judgment labels initially confusing, and P8 requested links to original claim sources; P7 nonetheless found the result *"easy to use, not bombarding with a lot of information,"* and P2 confirmed her earlier formative concerns (lack of interaction, vague nudges) had been addressed.

## C.5 Comparison with General-Purpose Tools

To situate Althea's outputs against tools a user might reach for instead, we selected the ten claims in Study 2's stimulus pool, chosen to span the pool's topical range (international relations, public health, domestic politics, celebrity and entertainment rumors) and its ground-truth distribution (five *Refuted*, two *Supported*, and three claims with *Conflicting Evidence* or mixed fact-checker verdicts), and submitted each to four general-purpose tools: Perplexity (Sonar, Perplexity's default model as of September 2025), ChatGPT (GPT-5), Google Search (organic results), and Google AI Mode (a Gemini-based model; Google does not publish exact version numbers for this feature). We ran this comparison only after data collection had concluded, specifically so that querying the external tools could not seed or bias the search indices, caches, or any of the models participants interacted with during the study.

Table 11. Post-hoc comparison of general-purpose tools against ten claims from Study 3's stimulus pool, run after data collection concluded. ✓ = verdict matched ground truth; × = verdict contradicted ground truth; ∼ = no clear verdict offered (see Section C.5 for how mixed-label claims and absent verdicts were scored).

| ID | Claim | Ground truth | Althea | Perplexity | ChatGPT | Google Search | Google AI Mode |
|---|---|---|---|---|---|---|---|
| STS001 | The United States of America and its Western allies have been using their media outlets to publish articles based on fabricated information under allegations of non-compliance with the Chemical Weapons Convention. | Refuted | ✓ | ✓ | ✓ | ✓ | ∼ |
| STS005 | Sleeping under a mosquito bed net treated with insecticide is ineffective and harmful to human health. | Refuted | ✓ | ✓ | ✓ | ✓ | ✓ |
| STS006 | Paul Pogba, who plays for Manchester United and the French national team, retired from international football in response to French President Macron's comments on Islamist terrorism. | Refuted | ✓ | ✓ | ✓ | ✓ | ✓ |
| STS014 | Scientific American magazine warned that 5G technology is not safe. | Conflicting | ✓ | ✓ | ✓ | ✓ | ✓ |
| STS015 | Duolingo apologized for calling JK Rowling 'mean' in a German lesson. | Supported | ✓ | ✓ | ✓ | ✓ | ✓ |
| STS016 | Obamas sold Martha's Vineyard house to former Epstein client Les Wexner. | Refuted | ✓ | ∼ | ✓ | ✓ | ✓ |
| STS017 | Trump ordered the clearance of homeless people nationwide. | Conflicting | ✓ | ✓ | ✓ | ∼ | ✓ |
| STS104 | A law called The Flora and Fauna Act classified aboriginal people as animals until Australian voters overturned it in the 1960s. | Refuted | ✓ | ✓ | ✓ | ✓ | ✓ |
| STS106 | China is pushing the frontiers of reproductive technology with the development of the world's first humanoid pregnancy robot. | Refuted | ✓ | × | ✓ | ∼ | ✓ |
| STS107 | Bill Gates-backed company created lab-made butter. | Supported | ✓ | ✓ | ✓ | ✓ | ✓ |
| **Correct / 10** | | | 10 | 8 | 10 | 8 | 9 |

Althea and ChatGPT both reached the correct verdict on all ten claims; Perplexity and Google Search each missed two, and Google AI Mode missed one (Table 11). Beyond accuracy, the comparison illuminates how a general-purpose tool's output, even a correct one, is a single free-text judgment the user must decide whether to trust, whereas Althea exposes the same underlying question, source credibility, expert consensus, corroborating evidence, and competing interpretations, as separable, inspectable steps. Althea's bottom-line verdict for every claim is in the table above; the rest of this section gives the full strategy-level trace behind one of them, STS016, as a worked example of what the Overall Judgment does not show on its own.

Table 12. *Althea*'s full strategy trace for STS016 ("Obamas sold Martha's Vineyard house to former Epstein client Les Wexner," *Refuted*), a worked example of the reasoning behind the single Overall Judgment reported for this claim in Table 11.

| Strategy | Output |
|---|---|
| Overall Judgment | **Refuted.** The Obamas never owned the Blue Heron Farm estate; they rented it for three summers beginning in 2009. The property was owned by Norman and Elena Foster, who sold it to a trust tied to Les Wexner in 2025. No evidence shows a direct sale from the Obamas to Wexner. |
| Source (credibility) | Traces the claim to a social media post: user-generated content on a platform with no unified editorial stance or verifiable author credentials, flagged as requiring independent verification rather than treated as a reported fact. |
| Fact-Checking (expert) | Surfaces a Snopes rating of *False* directly, with Snopes's own summary: the Obamas rented, not owned, the property; the 2025 sale was between the Fosters and Wexner. |
| Evidence (multi-hop) | Reconstructs the provenance of the sale (Foster ownership since 2011, the 2025 transaction via a trust managed by Wexner's attorney) and confirms the Obamas' role was limited to renting, at roughly $50,000 per week, during their vacations. |
| Controversial (perspectives) | Surfaces the claim's origin in conservative social media circles seeking to imply an Obama–Epstein connection via Wexner, and states plainly that this implication is not supported: Wexner's real association with Epstein is unrelated to any transaction involving the Obamas. |

For the same claim, ChatGPT, Google Search, and Google's AI Mode all matched the *Refuted* ground truth (Table 11). Perplexity is the instructive exception: its response opens by calling the claim "mostly true with some clarification," before concluding in its own final sentence that "it is false that the Obamas themselves sold the house to Wexner." We scored this as ambiguous rather than correct, since a user skimming only the headline judgment, and not the full response, would come away with the opposite of the fact-checkers' conclusion. This is the kind of case our own scaffolding is designed to surface and not resolve on the user's behalf: *Althea*'s Overall Judgment field states the verdict in the same sentence as the claim it is judging, and the four strategy panels above let a user see why Source, Fact-Checking, Evidence, and Controversial-perspective analysis converge on *Refuted*, instead of asking them to trust a single free-text summary that hedges internally.

## C.6 Analytical Results

The Cronbach alphas corresponding to the main measures used in the study are reported in Table 13.

## C.7 Claim accuracy

Table 14 reports the mean change in accuracy across intervention stages. Table 15 reports the mean change in confidence, while Table 16 reports the mean change in sharing inclinations across intervention stages. Table 17 reports the claims and their perceived accuracy across the different treatment conditions.

Table 13. Scale reliabilities and complete-case counts. News elaboration, Decision style, Familiarity, and CRT were administered only to the concurrently randomized phase-one sample ($N$ = 640); Self-search's instrument did not include them.

| Scale | $k$ | Cronbach's $\alpha$ | $N_{\text{complete}}$ |
|---|---|---|---|
| News elaboration | 6 | 0.808 | 640 |
| Decision style | 19 | 0.710 | 640 |
| Perception (Treatment conditions) | 5 | 0.726 | 422 |
| NASA–TLX | 5 | 0.731 | 422 |
| Engagement (Treatment conditions) | 12 | 0.877 | 422 |
| Familiarity | 12 | 0.894 | 640 |
| CRT | 3 | 0.719 | 640 |

Table 14. Mean change in accuracy across intervention stages, in percentage points. Values represent $M$ (SD).

| **Intervention Stage** | **Control** | **Summary Mode** | **Exploratory Mode** |
|---|---|---|---|
| After 1 Intervention | 0.36 (12.8) | 27.8 (23.8) | 26.6 (24.8) |
| After 2 Interventions | 0.46 (18.2) | 35.0 (33.7) | 33.3 (34.5) |
| After 5 Interventions | 1.82 (17.1) | 15.8 (27.4) | 16.2 (29.5) |
| Overall Improvement | 0.55 (7.13) | 16.7 (13.0) | 16.1 (14.1) |

Table 15. Mean change in confidence across intervention stages, expressed as a percentage of the 1–5 scale range (raw point change $\div$ 5 $\times$ 100; e.g., a full one-point increase registers as 20). Values represent $M$ ($SE$).

| **Intervention Stage** | **Control** | **Summary Mode** | **Exploratory Mode** |
|---|---|---|---|
| After 1 Intervention | 1.77 (0.68) | 19.7 (1.31) | 21.0 (1.22) |
| After 2 Interventions | 2.00 (0.78) | 20.0 (1.39) | 22.0 (1.37) |
| After 5 Interventions | 1.36 (1.09) | 20.8 (2.08) | 25.7 (1.97) |
| Overall Improvement | 1.89 (0.65) | 19.1 (1.21) | 20.3 (1.16) |

Table 16. Mean change in willingness to share across intervention stages. Negative values indicate reduced willingness to share.

| **Intervention Stage** | **Control** | **Summary Mode** | **Exploratory Mode** |
|---|---|---|---|
| After 1 Intervention | −0.14 (0.40) | −1.85 (0.83) | −2.06 (0.89) |
| After 2 Interventions | −0.02 (0.52) | −1.80 (0.92) | −1.24 (1.00) |
| After 5 Interventions | 1.04 (0.64) | 0.40 (1.40) | −0.43 (1.61) |
| Overall Improvement | −0.15 (0.37) | −1.93 (0.69) | −1.02 (0.70) |

### C.8 Auxiliary Analysis: System- and Individual-Level Heterogeneity

Alongside RQ1 and RQ2, we ran an auxiliary analysis of two classes of heterogeneity in Althea's effectiveness. *Individual differences* were assessed via linear regression with an interaction term between condition and partisan self-identification (7-point scale, recoded so higher values = more Republican); Table 18 reports the full covariate-adjusted model, and the interaction effects are further elaborated in Figure 9 and Appendix C.10. *System-level differences* (variance attributable to claim features and not participant characteristics) were assessed by examining claim-level accuracy gains, with each claim coded for veracity type (true, false, mixed/conflicting) and baseline difficulty (pre-task proportion correct); claims with low baseline accuracy represent cases where Althea's structured guidance should have the most room to improve performance, while claims with high baseline accuracy represent ceiling effects.

Table 17. Overview of experimental claims with post-task accuracy (% correct) by condition. Claims sorted by baseline difficulty (ascending). Baseline is pooled pre-task accuracy across conditions.

| ID | Veracity | Claim Text | Date | Baseline | Control | Summary | Exploratory |
|---|---|---|---|---|---|---|---|
| STS015 | True | Duolingo apologized for calling JK Rowling 'mean' in a German lesson. | 2025-08-20 | 37.9 | 50.0 | 93.0 | 90.1 |
| STS107 | True | Bill Gates-backed company created lab-made butter. | 2025-08-11 | 43.2 | 52.4 | 92.1 | 84.6 |
| STS006 | False | Paul Pogba, who plays for Manchester United and the French national team, retired from international football in response to French President Macron's comments on Islamist terrorism. | 2020-10-26 | 45.6 | 51.9 | 90.2 | 87.4 |
| STS104 | False | A law called The Flora and Fauna Act classified aboriginal people as animals until Australian voters overturned it in the 1960s. | 2020-10-11 | 46.6 | 53.4 | 88.1 | 90.3 |
| STS106 | False | China is pushing the frontiers of reproductive technology with the development of the world's first humanoid pregnancy robot. | 2025-08-15 | 50.1 | 54.6 | 62.7 | 72.3 |
| STS016 | False | Obamas sold Martha's Vineyard house to former Epstein client Les Wexner. | 2025-08-07 | 54.5 | 62.5 | 85.7 | 82.1 |
| STS001 | False | The United States of America and its Western allies have been using their media outlets to publish articles based on fabricated information under allegations of non-compliance with the Chemical Weapons Convention. | 2020-10-30 | 55.8 | 63.1 | 77.4 | 80.2 |
| STS017 | Mixed | Trump ordered the clearance of homeless people nationwide. | 2025-08-14 | 57.5 | 60.7 | 81.8 | 70.0 |
| STS005 | False | Sleeping under a mosquito bed net treated with insecticide is ineffective and harmful to human health. | 2020-10-26 | 60.0 | 69.2 | 82.1 | 87.0 |
| STS014 | Mixed | Scientific American magazine warned that 5G technology is not safe. | 2020-10-20 | 79.4 | 72.9 | 87.0 | 89.1 |

*Claim veracity and baseline difficulty.* The claim-level breakdown (Table 17) reveals systematic variance attributable to claim features. Baseline difficulty (pre-task accuracy) ranged from 37.9% (STS015) to 79.4% (STS014), confirming meaningful variation in the task's prior difficulty. Claims with the lowest baseline accuracy showed the largest absolute gains in both Althea conditions: STS015 improved from 37.9% to 93.0% in Summary and 90.1% in Exploratory; STS107 (baseline 43.2%) improved similarly. This floor-to-ceiling pattern is consistent with accuracy gains being limited by ceiling effects for already-easy claims and maximized for hard ones. Claim-level performance between the two Althea conditions is more evenly split than a single-exception framing would suggest: Summary exceeds Exploratory on five claims (STS015, STS107, STS006, STS016, STS017, including both true claims) and Exploratory exceeds Summary on the other five (STS104, STS106, STS001, STS005, STS014). The two largest gaps run in opposite directions: Exploratory's biggest margin is STS106 (72.3% vs. 62.7%, +9.6pp), Summary's is STS017 (81.8% vs. 70.0%, +11.8pp). We do not have a claim-content explanation for why any specific item favors one mode over the other, and would rather say so than retrofit one; the claim-level variance is real but its source is an open question. Together, these claim-level patterns indicate that system-side differences (specifically baseline difficulty and veracity type) are a real source of variance in Althea's effectiveness, over and above condition assignment.

## C.9 Attrition-Correction Detail: Naive vs. Corrected Wave-1 Baselines

Section 5.5.1 reports a difference-in-differences estimate built on each condition's attrition-corrected Wave-1 baseline (completers' own Wave-1 score, and not the full Wave-1 sample), and a sensitivity check showing the DiD conclusion holds either way. This appendix gives the full per-condition detail behind that correction: each condition's full Wave-1 sample compared against only the subset who returned for Wave 2, since a Wave-2 score does not exist for non-completers, alongside a paired test restricted to completers that compares their own Wave-1 and Wave-2 scores directly (Table 19).

Table 18. Linear regression of overall improvement (claims out of 5 improved from pre- to post-task) on treatment, covariates, and party identification interaction ($N$ = 640; 2 observations dropped for missing covariates). Reference category: Control condition.

| | Model 1 | | Model 2 | |
|---|---|---|---|---|
| **Predictor** | ***b*** **(SE)** | ***p*** | ***b*** **(SE)** | ***p*** |
| Intercept | 0.652 (0.218) | .003 ** | 0.663 (0.217) | .002 ** |
| Summary Mode | 1.393 (0.107) | <.001 *** | 1.376 (0.112) | <.001 *** |
| Exploratory Mode | 1.438 (0.109) | <.001 *** | 1.340 (0.117) | <.001 *** |
| Party identification | −0.068 (0.023) | .003 ** | −0.019 (0.037) | .600 |
| Cognitive reflection | −0.070 (0.040) | .081 . | −0.067 (0.040) | .097 . |
| Age (years) | −0.003 (0.003) | .293 | −0.003 (0.003) | .327 |
| Education (ordinal) | −0.074 (0.035) | .034 * | −0.075 (0.034) | .029 * |
| Income (ordinal) | −0.022 (0.014) | .118 | −0.020 (0.014) | .148 |
| Gender (ordinal) | 0.134 (0.088) | .128 | 0.135 (0.088) | .124 |
| Summary Mode × Party identification | — | — | −0.035 (0.055) | .527 |
| Exploratory Mode × Party identification | — | — | −0.123 (0.055) | .026 * |
| Residual SE | 1.11 | | 1.10 | |
| $R^2$ / Adjusted $R^2$ | .297 / .288 | | .303 / .292 | |
| $F$ (df) | 33.37 (8, 631)*** | | 27.35 (10, 629)*** | |

$^{\cdot}p < .10$; $^{*}p < .05$; $^{**}p < .01$; $^{***}p < .001$

Table 19. Attrition-robustness check: naive Wave-1 mean (all participants) vs. corrected Wave-1 mean (Wave-2 completers only), compared against each condition's Wave-2 mean in a paired test restricted to completers.

| Condition | $n$ completers | Naive W1 | Corrected W1 | W2 | Paired test |
|---|---|---|---|---|---|
| Control | 147 | 58.9% | 58.2% | 49.3% | $t(146) = 2.78, p = .006$ |
| Summary Mode | 153 | 84.0% | 85.2% | 57.5% | $t(152) = 8.92, p < .001$ |
| Exploratory Mode | 151 | 83.5% | 84.1% | 58.3% | $t(150) = 8.02, p < .001$ |
| Self-search | 147 | 73.2% | 75.8% | 77.1% | $t(146) = -0.48, p = .63$ |

For Control, Summary, and Exploratory, the correction changes almost nothing: the completers-only Wave-1 mean is within 1.5 percentage points of the full-sample mean for every condition, and the paired tests confirm significant decay in all three ($p = .006$, $p < .001$, $p < .001$ respectively). Self-search is the exception: its naive Wave-1 mean (73.2%) understates the completers' own Wave-1 accuracy (75.8%), and once compared to itself instead of to the full non-completing sample, Self-search shows no significant change over the ten-day interval in isolation (paired $t(146) = -0.48$, $p = .63$), in contrast to the increase the naive comparison suggests. This asymmetry is not a matter of applying scrutiny selectively: we ran the identical check on all four conditions, and Self-search is the only one where the correction changes the within-condition conclusion, consistent with its attrition rate being roughly double that of the other three. This per-condition correction is what feeds the Wave-1 baseline used in the main-text difference-in-differences estimate (Section 5.5.1).

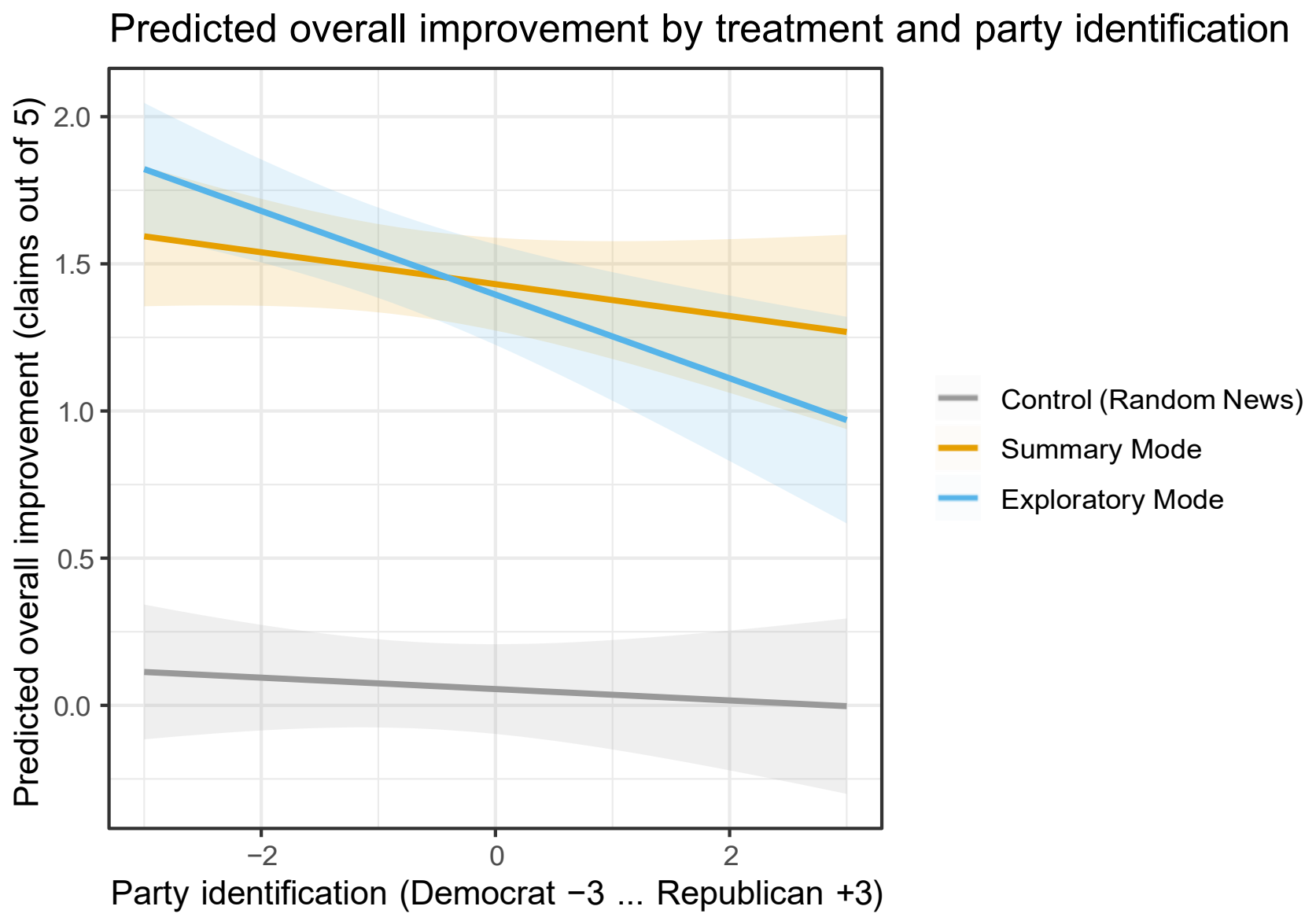

Fig. 9. Predicted overall improvement (claims out of 5) by treatment and party identification ($N = 640$). Lines represent estimated marginal means from the linear model, with shaded ribbons indicating 95% confidence intervals. Simple slopes: Control ($\beta = -0.019$, $p = .60$, n.s.), Summary Mode ($\beta = -0.054$, corrected $p = 1.00$, n.s.), Exploratory Mode ($\beta = -0.142$, corrected $p = .052$, n.s.; corrected for the two interaction terms tested in this model). The negative slope for Exploratory Mode is not a confirmed finding; see Appendix C.10.

*Self-search completers vs. non-completers.* Section 6.3 summarizes this check: Self-search's Wave-2 completers ($n = 147$, 45.8%) and non-completers ($n = 174$) are compared on Wave-1 baseline accuracy and demographics in Table 20. The two groups do not differ significantly on baseline accuracy, age, party identification, or gender; completers are modestly higher in education and income, both at $p < .05$ but neither survives a Bonferroni correction for the six characteristics tested here ($\alpha_{corrected} = .0083$). We read this as evidence against the specific worry that Self-search's Wave-2 sample is a subsample selected on prior fact-checking accuracy, while acknowledging a milder, SES-linked selection on who returns for a follow-up ten days later.

*Worst-case trim of the novel-item advantage.* Because Self-search's Wave-2 retention (45.8%) is lower than Control's, Summary's, and Exploratory's (67–75%), we additionally ask how much of Self-search's novel-item advantage (Section 5.5.1) could survive if its retained sample over-represents high performers by exactly the gap in retention rates. For each comparison condition, we compute the trim fraction $\beta = 1 - (s_{\text{Self-search}}/s_{\text{comparison}})$, remove the top $\beta$ share of Self-search's own observed novel-item scores (assuming, worst-case, that participants a higher retention rate would have additionally retained are drawn from the bottom of the remaining distribution), and recompute the mean on what remains (Table 21). The trimmed means (63–67%) still exceed every comparison condition's actual observed mean (58.7–61.4%), so the point estimate survives this worst-case reweighting; a one-sample $t$-test of each trimmed subsample against the corresponding comparison mean does not reach significance (all $p > .10$), but this is a conservative test by construction, since the trim simultaneously removes Self-search's highest scorers and roughly a third of its already-reduced completer sample ($n$ = 91–101 after trimming, versus 147 observed), leaving less power than the untrimmed comparison ever had. We read this as showing the advantage is not eliminated by a plausible worst-case reweighting, while stopping short of claiming that reweighted comparison is itself significant. It does not survive an even more extreme, and we think implausible, floor: if all 174 non-completers are assumed to have scored 0%, the population-weighted mean falls to 35.3%; solving for the break-even point, non-completers would need to have averaged 43–48% (below every comparison condition's own observed mean) for Self-search's advantage to disappear entirely.

Table 20. Self-search Wave-2 completers ($n$ = 147) vs. non-completers ($n$ = 174) on Wave-1 characteristics.

| Characteristic | Completers | Non-completers | Test |
|---|---|---|---|
| W1 accuracy (%) | 75.8 | 71.0 | $t(319) = -1.43, p = .154$ |
| Age (years) | 41.4 | 40.8 | $t(314) = -0.35, p = .728$ |
| Education (ordinal) | 4.45 | 4.14 | $t(313) = -2.15, p = .033$ |
| Income (ordinal) | 5.43 | 4.71 | $t(303) = -2.00, p = .046$ |
| Party scale | −0.69 | −0.58 | $t(307) = 0.53, p = .594$ |
| % Female | 55.8 | 60.3 | $\chi^2(2) = 0.68, p = .710$ |

Table 21. Worst-case sensitivity trim of Self-search's Wave-2 novel-item accuracy ($n$ = 147 completers, observed mean 77.1%), matching each comparison condition's higher Wave-2 retention rate under a monotonic-selection assumption (the participants a lower retention rate would have additionally captured are assumed to be the lowest-scoring remaining ones).

| Comparison | Retention | Trim | Trimmed mean | Actual mean |
|---|---|---|---|---|
| vs. Control | 67.3% | 32.0% | 66.7% | 61.4% |
| vs. Summary | 70.6% | 35.1% | 64.9% | 60.8% |
| vs. Exploratory | 74.5% | 38.5% | 63.0% | 58.7% |

## C.10 Individual-Difference Analysis: Partisan Moderation, Full Detail

This appendix reports, in full, an auxiliary individual-level analysis: a partisan interaction in Exploratory Mode that did not survive correction for multiple comparisons.

In the *Control* condition, partisan orientation was unrelated to accuracy (simple slope $\beta = -0.019$, 95% CI [−0.092, 0.053], $p = .60$; not part of the correction family below, since it is the reference-category main effect and not an interaction term). The two treatment×party interaction terms estimated in the regression (Summary×party, Exploratory×party; Table 18) constitute a natural family for correction, since both are tested in the same model. After Bonferroni correction for these two tests, neither is significant: *Exploratory Mode* (simple slope $\beta = -0.142$, 95% CI [−0.222, −0.062], corrected $p = .052$), where Democrat-leaning participants achieved $M$ = 85.3% accuracy versus $M$ = 78.8% for Republican-leaning participants, and *Summary Mode* (simple slope $\beta = -0.054$, 95% CI [−0.134, 0.026], corrected $p = 1.00$), where Democrat-and Republican-leaning participants achieved closer accuracy ($M$ = 84.1% vs. $M$ = 83.0%). Holm and Benjamini-Hochberg corrections give the same conclusion for both terms. Exploratory's corrected $p$-value falls just short of $\alpha$ = .05.

We tested age and gender as moderators in the identical model specification (treatment×moderator, same covariates) and found no comparable pattern: both interaction terms were far from significant for age ($p$ = .20, $p$ = .71) and for gender ($p$ = .78, $p$ = .65), uncorrected, indicating the partisan trend is not one hit among several demographic moderators that would emerge if tested indiscriminately.

### C.11 Mediation Test: Strategy Breadth Between Condition and Wave-2 Accuracy

Section 6.1 claims that Exploratory's behavioral persistence and its accuracy persistence dissociate: it retains more strategy-use breadth than Summary at Wave 2, but not a Wave-2 accuracy advantage. We tested this formally with a mediation model among Wave-2 completers in the two logged conditions ($n$ = 303): condition (Exploratory vs. Summary) predicting Wave-1 strategy breadth (path $a$), and breadth predicting Wave-2 accuracy controlling for condition and Wave-1 baseline accuracy (path $b$), with 2,000 nonparametric bootstrap simulations for the indirect effect.

Table 22. Mediation of the condition → Wave-2 accuracy relationship by Wave-1 strategy breadth, Exploratory vs. Summary, Wave-2 completers only ($N$ = 303). Path $b$ and the total/direct effect models control for Wave-1 baseline accuracy and covariates (cognitive reflection, age, education, income, gender), omitted here for space.

| Quantity | Estimate | 95% CI | p |
|---|---|---|---|
| Path $a$: Condition → Breadth (W1) | 1.074 | [0.72, 1.43] | <.001 *** |
| Path $b$: Breadth (W1) → Accuracy (W2) | −0.520 | [−3.31, 2.27] | .716 |
| Total effect: Condition → Accuracy (W2) | 0.703 | [−7.79, 9.13] | .868 |
| Direct effect (ADE) | 1.261 | [−7.46, 10.15] | .803 |
| Indirect effect (ACME) | −0.558 | [−3.75, 2.69] | .728 |
| Proportion mediated | −0.793 | [−6.46, 5.46] | .960 |

Path $a$ confirms what Section 5.5.2 already reports: Exploratory participants use significantly more distinct strategies than Summary participants ($b$ = 1.07, $SE$ = 0.18, $p$ < .001). Path $b$ is null: strategy breadth does not predict Wave-2 accuracy once Wave-1 baseline is controlled ($b$ = −0.52, $SE$ = 1.42, $p$ = .72). The total effect of condition on Wave-2 accuracy is itself null ($b$ = 0.71, $p$ = .87), consistent with Exploratory and Summary being statistically indistinguishable at Wave 2 (Section 5.5.2), so there is little total effect available for breadth to mediate in the first place. The indirect effect (ACME) is correspondingly null (−0.56, 95% CI [−3.75, 2.69], $p$ = .73), as is the direct effect (ADE = 1.26, $p$ = .80). We read this as a positive, quantitative confirmation of the dissociation already claimed qualitatively: Exploratory's behavioral gains and its (absent) accuracy advantage are separate phenomena, and one does not cause the other.

### C.12 Entry-Order Analysis: First-Strategy Choice, Breadth, and Accuracy

Section 5.5.2 reports which strategies participants used; this section asks which strategy participants reached for first, and whether that choice matters for what follows. For each participant in the two logged conditions, we identified the first user-selectable strategy (Source Analysis, Expert Search, Perspective Integration, or Multi-Hop Reasoning; Evidence Synthesis is a system-side step and is excluded, Section 5.5.2) invoked in their opening conversation, timestamped from the interaction logs. An identifiable first strategy was recoverable for 431 of 557 logged-condition participants (77.4%);

of these, 339 (179 Exploratory, 160 Summary) had matched Wave-1 breadth and accuracy data and form the analytic sample below.

Table 23. Distribution of participants' first-selected strategy in their opening conversation, by condition, with mean Wave-1 strategy breadth ($n_strategies_distinct$, of five) among participants for whom breadth and accuracy data were also available ($N$ = 339: 179 Exploratory, 160 Summary).

| | Exploratory | | Summary | |
|---|---|---|---|---|
| **First Strategy Selected** | **n (%)** | **Mean Breadth** | **n (%)** | **Mean Breadth** |
| Source Analysis | 128 (71.5%) | 4.65 | 114 (71.3%) | 4.28 |
| Expert Search | 39 (21.8%) | 4.49 | 29 (18.1%) | 3.69 |
| Multi-Hop Reasoning | 11 (6.1%) | 4.18 | 12 (7.5%) | 3.17 |
| Perspective Integration | 1 (0.6%) | 4.00 | 5 (3.1%) | 3.80 |

Source Analysis is the dominant entry point in both conditions: 71–73% of participants open with it, more than three times the next most common choice (Expert Search, 18–22%), with Multi-Hop Reasoning and Perspective Integration a distant third and fourth in both conditions (Table 23). This ordering matches the module's position first in the interface sidebar at least as plausibly as it reflects a reasoning preference, and we do not read it as evidence of a spontaneous strategic preference.

Entry choice does, however, predict what happens afterward. In a model of Wave-1 strategy breadth on entry strategy and condition, entry strategy has a significant effect ($F(3, 334) = 6.82$, $p < .001$) net of condition ($F(1, 334) = 21.70$, $p < .001$): participants who started with Source Analysis or Expert Search went on to use more distinct strategies than those who started with Multi-Hop Reasoning, in both conditions (Table 23). Entry choice does not predict Wave-2 accuracy improvement in the equivalent model ($F$ (3, 334) = 0.62, $p$ = .60; condition $F$ (1, 334) = 2.93, $p$ = .09). We read this as consistent with the dissociation reported in Appendix C.11: process variables that track engagement breadth, whether condition itself or which strategy a participant reaches for first, predict further engagement but not accuracy. Which strategy a participant starts with is a weak causal claim in an observational log (participants were not randomized to entry strategy), so we treat this as descriptive of engagement patterns and not a mechanism result.